# Multipath Multiplexing for Capacity Enhancement in SIMO Wireless Systems (Draft with more Results and Details)

Tadilo Endeshaw Bogale, *Member, IEEE*, Long Bao Le, *Senior Member, IEEE*, Xianbin Wang *Fellow, IEEE* and Luc Vandendorpe *Fellow, IEEE*

## Abstract

This paper proposes a novel and simple orthogonal faster than Nyquist (OFTN) data transmission and detection approach for a single input multiple output (SIMO) system. It is assumed that the signal having a bandwidth $B$ is transmitted through a wireless channel with $L$ multipath components. Under this assumption, the current paper provides a novel and simple OFTN transmission and symbol-by-symbol detection approach that exploits the multiplexing gain obtained by the multipath characteristic of wideband wireless channels. It is shown that the proposed design can achieve a higher transmission rate than the existing one (i.e., orthogonal frequency division multiplexing (OFDM)). Furthermore, the achievable rate gap between the proposed approach and that of the OFDM increases as the number of receiver antennas increases for fixed $L$. This implies that the performance gain of the proposed approach can be very significant for a large-scale multi-antenna wireless system. The superiority of the proposed approach has been shown theoretically and confirmed via numerical simulations. Specifically, we have found upper-bound average rates of 15 bps/Hz and 28 bps/Hz with the OFDM and proposed approaches, respectively, in a Rayleigh fading environment with 32 receive antennas and signal to noise ratio (SNR) of 15.3 dB. The extension of the proposed approach for different system setups and associated research problems is also discussed.

## Index Terms

I. INTRODUCTION

The study in Wireless World Research Forum (WWRF) has predicted that 7 trillion wireless devices will be supported by wireless networks, both for human and machine-type communications, in 2017. Furthermore, the mobile data traffic has increased dramatically over the years, which is mainly driven by the massive demand of data-hungry devices such as smart phones, tablets and broadband wireless applications such as multimedia, 3D video games, e-Health, Car2X communications [1], [2]. It is expected that the future 5G wireless network will deliver about 1000 times more capacity than that of the current 4G system. To meet this demand different enabling techniques have been under development including massive multiple input multiple output (MIMO), millimeter wave, full duplex communication and faster than Nyquist (FTN) transmission [3]–[6].

The first work on FTN transmission was proposed by Mazo in 1975 where it is claimed that it is possible to achieve a transmission rate beyond what can be achieved with the Nyquist criterion without requiring additional bandwidth [7]. In general, the symbol error probability plays a fundamental role in understanding a communication strategy which can be directly related to the minimum Euclidean distance $d_{min}$ between two distinct symbols. The key idea of [7] is that for the scenario with Nyquist symbol duration $T_{Ny}$ and sinc pulse shaping filter, $d_{min}$ remains the same even when we pack a symbol in $\tau T_{Ny}$ with $\tau \in [0.802, 1]$. This result implies that one can still maintain the same error probability as that of the Nyquist transmission by selecting $\tau \approx 0.82$ which in turn helps to improve the spectrum efficiency by $\frac{1}{0.802} \approx 25\%$. It is also shown that different pulse shaping filters result in different Mazo limits (i.e., minimum $\tau$). For instance, the minimum $\tau$ with a root raised cosine filter with excess bandwidth $\beta = 0.3$ yields a 42% increase in bandwidth efficiency (i.e., $\tau_{min} = 0.703$) [8].

These results develop new perspectives about the achievable rate and have recently attracted researchers (see for example [3], [4], [9]). Having said this, however, there is one major difference between Nyquist and FTN transmissions. The data symbols transmitted with the Nyquist approach can be decoded independently on a symbol-by-symbol basis with appropriate matched filtering which consequently facilitates low complexity receiver. However, as the FTN transmission does not enable orthogonal symbol-by-symbol detection, it requires more complicated symbol detections [3], [4], [10]. In this regard, low-complexity receiver algorithms are suggested by several researchers such as a successive interference cancellation approach in [11] (see also [12], [13] and their references). A detailed survey of FTN transmission for different system parameters and its implementation aspects can be found in [4] where it is stated that such a transmission approach can provide up to twice the spectrum efficiency compared to the existing Nyquist transmission (i.e., 2 times the rate achieved by the single input single output (SISO) system).

The FTN transmission can be realized both for single-antenna and multi-antenna systems [14]. It is shown that for the single input multiple output (SIMO) and MIMO channels, the FTN transmission also improves the performance of the system [15]. In fact, the FTN allows the transmitter to send data symbols at a rate higher than the Nyquist transmission which can also be interpreted as a novel approach to exploit the additional degrees of freedom of the channel (i.e., "multiplexing gain"). This recalls us the two fundamental aspects of a wireless channel (i.e., diversity and multiplexing) which were studied in the existing MIMO communication literature [16], [17]. The multi-antenna system can be designed as a SIMO, MIMO and MISO (multiple input single output). It is well known that when either the transmitter or receiver has single antenna (i.e., MISO and SIMO), the communication system can achieve only a diversity (power) gain [16]–[18].

For MIMO systems, one can achieve both diversity and multiplexing gains under favorable channel conditions (i.e., it is possible to transmit and reliably detect on a symbol-by-symbol basis more than one symbol per Nyquist period). On the other hand, it has been shown in [16] that there is a fundamental tradeoff between the diversity and multiplexing gains in a MIMO system (i.e., one can increase diversity (multiplexing) gain while decreasing multiplexing (diversity) gain). A detailed analysis of the MIMO system performance for different parameter setups and signal to interference plus noise ratio (SINR) levels for flat fading channels can be found in various wireless communication books (see Chapters 5-7 of [16] for extensive study).

The design of efficient transmission strategies for flat fading channels can be extended almost straightforwardly to the frequency selective channels. This is due to the fact that any signal transmitted through a frequency selective channel can be splitted into orthogonal chunks of signals (i.e., sub-carriers) each experiencing a flat fading channel by using the orthogonal frequency division multiplexing (OFDM) transmission approach. By employing the OFDM scheme, the

achievable rate per Hertz of a SIMO system can be expressed as [16]

$$R_{Ex} = \frac{1}{N_s} \sum_{i=1}^{N_s} \log_2(1 + SNR_{Ex}^i(N)) \text{ bps/Hz} \quad (1)$$

where $N_s$ is the number of sub-carriers which increases with the transmission bandwidth $B$ Hz for fixed sub-carrier spacing, $SNR_{Ex}^i(N)$ is the signal to noise ratio (SNR) achieved at sub-carrier $i$ of the OFDM transmission approach, which is a function of the number of receive antennas $N$. The SNR achieved at each sub-carrier depends on different factors ranging from the number of antennas to the channel condition. In a multi-antenna system, $R_{Ex}$ is usually quantified by assuming that the channel between any transmit receive pair experiences independent fading [16]. Also, the deployment of large antenna arrays at the base station (BS) has been prototyped and studied where recent measurement results [19]–[24] suggest that despite the statistical difference between the measured channels and the independent and identically distributed (i.i.d) channels, most of the theoretical conclusions made under the independence assumption are still valid for the realistic massive MIMO channels[1]. Now when the channel between the transmitter and each receiver has $P$ unit-variance independent multipath components, the existing OFDM transmission can have the following upper-bound average rate [17][2]

$$R_{Ex} = \log_2\left(1 + \frac{NP}{\sigma^2}\right) \quad (2)$$

where $\sigma^2$ is the noise variance. As we can see from this expression, increasing the number of receiver antennas $N$ enables to enhance the channel achievable rate (2) *logarithmically* only [17], [26].

**Motivation and Objective**: The FTN transmission can be realized for different systems such as massive MIMO, mmWave and full-duplex wireless systems. Thus, such a transmission strategy will undoubtedly improve the overall capacity of the future 5G network. However, the existing FTN transmission can improve the overall point-to-point achievable rate at most two times only[3]. For instance, if we consider a SIMO system, the capacity of FTN transmission will be at most $2R_{Ex}$. Furthermore, the receiver may need to use a non-linear (joint) decoding approach which is often not desirable in practice. This motivates us to examine the following problem: Is it possible to employ a FTN transmission approach while ensuring that the point-to-point link achievable rate is improved more than 2 times with a simple symbol-by-symbol decoding? If yes, in what scenario and how can we enable the symbol-by-symbol decoding?

---

[1]This is also stated as a "favorable propagation condition" which is commonly adopted in MIMO channels [16], [17].

[2]As can be seen in (1), the accurate rate expression of the $i$th sub-carrier is $R_i = \log_2\left(1 + \frac{|\mathbf{h}_{oi}|^2}{\sigma^2}\right)$, and the true average value of $R_i$ is given as $R_i^{av} = \text{E}\{R_i\}$ where $\mathbf{h}_{oi}$ is the channel between the transmitter and receiver of sub-carrier $i$. Since the exact analytical expression of $R_i^{av}$ is difficult to derive for most practically relevant statistical models of $\mathbf{h}_{oi}$ (for example, i.i.d case), we use the well known Jensen's inequality to get the upper-bound expression for $R_i^{av}$ by replacing $|\mathbf{h}_{oi}|^2$ with its average value [25].

[3]This can be translated to a maximum of 2 times system capacity compared to the Nyquist transmission.

In the current paper, we propose a novel FTN transmission and symbol-by-symbol detection approach for SIMO systems over wideband channels (i.e., frequency selective fading channels). The proposed transmission scheme exploits the inherent multipath characteristics of wideband channels. We show that when the channel has $P$ independent multipath components (i.e., a favorable propagation condition) which is commonly adopted in different MIMO channels [16], [17] and the receiver has multiple antennas $N \geq P$, one can achieve the following upper-bound rate[4]

$$R_{Pr} = \frac{P^2}{3P-2} \log_2\left(1 + \frac{3P-2}{P^2}\left(\frac{N-P-1}{\sigma^2}\right)\right). \tag{3}$$

For fixed $P$ and $\sigma^2$, one can notice that $R_{Gap} = R_{Pr} - R_{Ex}$ increases as $N$ increases which renders the proposed approach appealing for large receiver antenna array systems. The complexity of the proposed approach is almost the same as that of the OFDM transmission approach and it is simple to implement. As will be clear in Section IV, the proposed design allows the estimated symbols to be decoded independently without experiencing inter-symbol-interference (ISI). For this reason, one can consider the proposed transmission strategy as orthogonal faster than Nyquist (OFTN). The rate expressions $R_{Pr}$ and $R_{Ex}$ have been given for correlated multipath channel components as well. The proposed approach also does not suffer from the peak to average power ratio (PAPR) problem which provides additional advantage compared to OFDM. On the other hand, one can straightforwardly extend the proposed transmission approach to multiuser (massive) MIMO systems.

We would like to emphasize here that the proposed approach relies solely on the number and characteristics of multipath components, and number of antennas at the receiver side. Moreover, the proposed OFTN transmission exploiting the multiplexing gain which is incorporated in multipath channels has never been revealed in the existing works (to the best of our knowledge).

---

[4]As will be clear in the simulation section, $R_{Pr}$ and $R_{Ex}$ are almost the same as their true average values for practically relevant $N$ and $P$.

This motivates us to use the term *multipath multiplexed* transmission. Indeed, different possible bandwidths are utilized in the current long term evolution (LTE) network (see [27], [28] for more details). For instance, if we consider a $B = 2.5$ MHz bandwidth, the number of multipath components is around 18 which is significant [29]. Increasing $B$ would increase the number of multipath components in a given transmission environment.

This paper is organized as follows. Sections II and III discuss the considered signal and channel models, and a brief summary of the OFDM and problem formulation of the proposed design, respectively. The detailed descriptions of the proposed OFTN transmission and detection approach, and its performance analysis are presented in Sections IV and V, respectively. The effect of power adaptation on the performance of OFDM and OFTN designs, and some of the practical issues of the OFTN scheme including its similarities with that of FTN is explained in Sections VI - VII. Simulation results are provided in Section VIII. Finally, some open research problems and conclusions are given in Sections IX and X, respectively.

*Notations:* In this paper, upper (lower-case) boldface letters denote matrices (column) vectors. $\mathbf{X}_{(i,j)}$, $\mathbf{X}^T$, $\mathbf{X}^H$ and $\mathrm{E}(\mathbf{X})$ denote the $(i,j)$th element, transpose, conjugate transpose and expected value of $\mathbf{X}$, respectively. The $\mathrm{convmtx}(.)$, $\mathbf{1}_N$, $\mathbf{I}$ and $\mathbb{C}^{N \times M}(\Re^{N \times M})$ denote convolution matrix, an $N$ sized vector of ones, identity matrix and $N \times M$ complex (real) entries, respectively.

## II. SIGNAL AND CHANNEL MODEL

We consider a single user system where the transmitter has one antenna whereas, the receiver has $N \gg 1$ antennas. The transmitted signal has a transmission bandwidth $B$ which is sufficiently large making the channel to be frequency selective (i.e., wideband signal). The channel impulse response between the transmitter and the $n$th receive antenna is denoted by $h_n(t)$ which is assumed to be linear time invariant. In the following, we provide the discrete time baseband equivalent of $h_n(t)$. For better exposition, we first consider a SISO channel (i.e., ignoring the subscript $n$) and the result is extended further to a SIMO channel.

TABLE I: Frequently used variables, vectors and matrices

| Var. | Definition |
|---|---|
| $\mathbf{Q}$ | Spatial channel correlation matrix |
| $\mathbf{R}$ | Temporal channel correlation matrix |
| $N$ | Number of receive antenna |
| $S$ | Number of transmitted symbols in 3P-2 sampling period |
| $P$ | Number of multipath components of the channel impulse response |
| $L$ | Number of multipath taps in OFDM transmission ($L = P$ when $g(t) = \text{rect}(t)$) |
| $g(t)$ | Transmitter and receiver pulse shaping filter in existing OFDM (i.e., $g(t) = \text{rect}(t)$ of width $T_s$ |
| $\tilde{g}(t)$ | Transmitter and receiver pulse shaping filter of proposed design (i.e., $\tilde{g}(t) = \text{rect}(t)$ of width $L\tilde{T}_s$ |
| $Q$ | Width of the transmitter and receiver pulse shaping filters ($Q = L = P$ when $\tilde{g}(t) = \text{rect}(t)$) |
| $B$ | Signal bandwidth ($B = \frac{1}{T_s}$) |
| $T_d$ | Channel delay spread |
| $\tilde{T}_s$ | Temporal spacing of equally spaced independent channel coefficients ($\tilde{T}_s = \frac{T_d}{P}$) |
| $T_b$ | Block duration of OFDM |
| $N_s$ | Number of sub-carriers of OFDM (it is assumed that all sub-carriers are used) |
| $s[k]$ | Transmitted data signal in $k$th symbol period of OFDM |
| $d_i, \forall i$ | Transmitted symbols of proposed approach which are temporally separated by $\tilde{T}_s$ |
| $y_l$ | Received sample at time slot $l$ obtained at each antenna (i.e., (24)) |
| $\bar{\mathbf{H}}_I$ | Equivalent random channel between transmitter and all receivers when $y_l$ of all antennas are stacked |
| $\bar{\bar{\mathbf{H}}}$ | Overall channel between transmitter and all receivers when $y_l$ of all antennas are stacked $\bar{\bar{\mathbf{H}}} = \mathbf{Q}^{1/2}\bar{\mathbf{H}}_I\mathbf{R}^{1/2}$ |
| $h_i, \forall i$ | Discrete time channel impulse response coefficients with temporal spacig $\tilde{T}_s$ |

If we transmit an arbitrary baseband waveform $x(t)$ having bandwidth $B$ and carrier frequency $f_c$, the received signal in baseband form can be expressed as [17], [30]

$$y(t) = h(t) * x(t) + w(t) \tag{4}$$

where $h(t)$ is the baseband equivalent channel impulse response, $w(t)$ is the additive noise and $(*)$ is the convolution operator. In general, one can model $h(t)$ as (see (3.10) of [30], (2.19) of [17] and [31])

$$h(\tau) = \sum_{i=0}^{P-1} \alpha_i e^{j\phi_i(\tau)} \delta(\tau - \tau_i) \tag{5}$$

where $P$ is the total number of multipath components, and $\alpha_i$, $\phi_i$ and $\tau_i$ are the attenuation (due to pathloss and shadowing which is varied quite slowly), phase and propagation delay of the channel from the transmitter to the receiver in path $i$, respectively. Here we assume, without loss of generality, that each multipath component corresponds to a single reflector/scatterer and $\phi_i(\tau) = 2\pi f_c \tau_i + \phi_{f_{D_i}}(\tau)$ with $\phi_{f_{D_i}}(\tau) = 2\pi\tau f_{D_i}$, $f_{D_i} = \frac{v}{\lambda}\cos\theta_i$ (i.e., Doppler shift) and $\theta_i$ is the angle of arrival (AOA) of path $i$. As can be seen from the above expression, a small change in $\tau_i$ can lead to a significant variation in the phase $\phi_i$. Thus, the rate of change of $\phi_i$ can be in the order of very few milliseconds [17], [30].

TABLE II: LTE Extended Typical Urban model (ETU)

| Delay (ns) | 0 | 50 | 120 | 200 | 230 | 500 | 1600 | 2300 | 5000 |
|---|---|---|---|---|---|---|---|---|---|
| Relative PDP (dB) | -1 | -1 | -1 | 0 | 0 | 0 | -3 | -5 | -7 |

In general, the channel impulse response (5) is a random process due to the random amplitude, phase and delay components. Thus, $h(t)$ can be characterized only statistically or from experimental results. In a practical setup, the channel response corresponding to $\tau_i$ is uncorrelated with that of $\tau_j, j \neq i$. This is mainly due to the fact that these two channel responses are the results of different scatterers (i.e., uncorrelated scattering) [30] (see page 181). Furthermore, as $f_c\tau_i = \frac{d_i}{\lambda}$ with $d_i \gg \lambda$, where $d_i$ is the propagation distance of the $i$th path, $\phi_i$ can be reasonably modeled as a uniformly distributed random variable $\mathcal{U}[0,\ 2\pi]$ [17], [30]. The power delay profile (PDP) of a multipath channel can therefore be given as

$$A_h(\tau_i, \tau_j) = \mathrm{E}\{h(\tau_i)h(\tau_j)^H\} = \mathrm{E}\{\alpha_i\alpha_j\}\delta|i-j| \triangleq A_h(\tau_i) = \alpha_i^2. \quad (6)$$

In a wireless channel, the delay spread $T_d$ is defined as the time index $\tau_i$ where $A_h(\tau_i) \approx 0^5$. Different standards suggest different delay parameters and PDP's for different environments. For instance, the Extended Typical Urban model (ETU) of the LTE network uses the specification given in Table II [32]. As can be seen from this table, the LTE ETU channel model contains $P = 9$ independent paths where each of them has a different PDP (see also [33] for other channel models). Furthermore, one can observe that the time delays may not be spaced uniformly.

Thus, without loss of generality, one can represent the discrete time baseband equivalent form of $h(t)$ as

$$\mathbf{h} = [h_1, h_2, \cdots, h_P]^T = \mathbf{R}^{1/2}\bar{\mathbf{h}} \quad (7)$$

where $\bar{\mathbf{h}} \in \mathcal{C}^{P\times 1}$ is a vector containing independent channel coefficients (i.e., as in Table II), $\mathbf{R} \in \mathcal{C}^{P\times P}$ is a temporal correlation matrix and $h_1, \cdots, h_P$ are equally spaced channel coefficients with temporal spacing given by $\tilde{T}_s \triangleq \frac{T_d}{P}$ (for example $T_d = 5\mu s$ and $P = 9$ if we apply the ETU channel model given in Table II). From these explanations, one can understand that $T_d$ is determined by the effects of reflector and scatterers which are mainly related to the propagation environments. Specifically, one may expect to experience very large $T_d$ in a densely populated urban areas, and very small value of $T_d$ in an indoor environment [17], [27], [28], [30], [34]. The coherence bandwidth which explains how fast the channel changes in frequency is given by

$$W_c = \frac{1}{T_d}. \tag{8}$$

---

[5]The multipath delay spread $T_d$ can also be defined as the difference in the propagation delay between the longest and shortest path having a significant energy, i.e., $T_d = \max_{i,j}|\tau_i - \tau_j|$.

The channel coherence bandwidth $W_c$ can also be treated as the bandwidth in which the channel is considered as 'almost constant'. Now when $B \gg W_c$, the wireless channel is termed as a frequency selective channel (or wideband channel) which is the focus of the current paper. When the receiver has $N$ antennas, one can model the channel coefficient between the transmitter and each receiver with (7).

## III. EXISTING OFDM TRANSMISSION AND PROBLEM STATEMENT

This section provides a brief summary of the well known OFDM transmission and formally states the considered problem.

### A. Summary of OFDM Transmission

When the channel is frequency selective, OFDM is known to be an efficient transmission strategy to maximize the achievable rate [17]. In fact, OFDM can be used to harness the effect of ISI that the transmitted signal experiences when $B \gg W_c$. For brevity, this section briefly summarizes the main idea of OFDM transmission and reception. In the existing transmission approach, the baseband transmitted signal $x(t)$ can be expressed as

$$x(t) = \sum_k s[k]g(Bt - k) \tag{9}$$

where $s[k], \forall k$ are the transmitted symbols with period $T_s = \frac{1}{B}$ and $g(t)$ is the pulse shaping filter. In most wireless systems, $g(t)$ is designed as a rectangular, raised cosine or sinc function.

In this section, we also start with a SISO system (i.e., by dropping the antenna subscript $n$) and then extend the analysis for SIMO system. When the receiver has single antenna, the received

signal $y(t)$ can be expressed as

$$y(t) = h(t) * x(t) + w(t) = \sum_k s[k] \sum_{i=0}^{P-1} h_i g(B(t - i\tilde{T}_s) - k) + w(t). \tag{10}$$

The recovered signal at the $m$th symbol period $y[m] = y(mT_s)$ can be rewritten as (see (2.35) of [17])

$$y[m] = \sum_{l=0}^{L-1} f_l s[m-l] + w[m] = \sum_{l=0}^{L-1} \tilde{h}_{L-1-l} s[m-l] + w[m] \tag{11}$$

where $w[m]$ is the noise sample and

$$f_l = \sum_{i=0}^{P-1} h_i g\left(l - \frac{\tilde{T}_s}{T_s} i\right), \quad [\tilde{h}_0, \tilde{h}_1, \cdots, \tilde{h}_{L-1}] = [f_{L-1}, f_{L-2}, \cdots, f_0]. \tag{12}$$

As can be seen from (11), $y[m]$ contains not only $s[m]$ but also additional interference terms $s[m-l], \forall l$, where $l$ depends on the channel and the support width of $g(t)$. If we consider $g(t) = \text{rect}(t)$ (i.e., $g(t) = 1, t \in [0, T_s]$ and $g(t) = 0, t > T_s$), we will have $L = \frac{\tilde{T}_s P}{T_s} = \frac{T_d}{T_s} = P$ (i.e., the number of multipath taps) [35]. Now if the OFDM signal has a block length $T_b = N_s T_s$, the received samples $y[1], y[2], y[3], \cdots, y[N_s]$ can be stacked in vector form $\mathbf{y}$ which can be expressed as

$$\mathbf{y} = \tilde{\mathbf{H}} \mathbf{s} + \mathbf{w} \tag{13}$$

where $\mathbf{s} = [s[1], s[2], \cdots, s[N_s]]$, $\mathbf{w} = [w[1], w[2], \cdots, w[N_s]]$ and $\tilde{\mathbf{H}}$ is a Toeplitz matrix formed by using the channel coefficients $[\tilde{h}_0, \tilde{h}_1, \cdots, \tilde{h}_{L-1}]$ [17], [18], i.e.,

$$\tilde{\mathbf{H}} = \begin{bmatrix} \tilde{h}_{L-1} & 0 & 0 & 0 & 0 & 0 & 0 & 0 & 0 & 0 \\ \tilde{h}_{L-2} & \tilde{h}_{L-1} & 0 & 0 & 0 & 0 & 0 & 0 & 0 & 0 \\ \vdots & \vdots & \vdots & 0 & 0 & 0 & 0 & 0 & 0 & 0 \\ \tilde{h}_0 & \tilde{h}_1 & \tilde{h}_3 & \cdots & \tilde{h}_{L-2} & \tilde{h}_{L-1} & 0 & 0 & 0 & 0 \\ 0 & \tilde{h}_0 & \tilde{h}_1 & \tilde{h}_3 & \cdots & \tilde{h}_{L-2} & \tilde{h}_{L-1} & 0 & 0 & 0 \\ 0 & 0 & \tilde{h}_0 & \tilde{h}_1 & \tilde{h}_3 & \cdots & \tilde{h}_{L-2} & \tilde{h}_{L-1} & 0 & 0 \\ 0 & 0 & 0 & \tilde{h}_0 & \tilde{h}_1 & \tilde{h}_3 & \cdots & \tilde{h}_{L-2} & \tilde{h}_{L-1} & 0 \\ \vdots & & & & \vdots & & & & & \vdots \\ 0 & 0 & \cdots & 0 & \tilde{h}_0 & \tilde{h}_1 & \tilde{h}_3 & \cdots & \tilde{h}_{L-2} & \tilde{h}_{L-1} \end{bmatrix}. \tag{14}$$

Since $\tilde{\mathbf{H}}$ is a full-rank matrix, the receiver can reliably estimate $\mathbf{s}$ by utilizing appropriate equalization techniques (for example with a zero forcing (ZF) approach which is a simple channel inversion technique). However, such an approach creates undesirable effects such as noise enhancement.

The existing OFDM transmission and reception approach aims to circumvent this drawback, and if we append the last $L$ symbols of $\mathbf{s}$ in the beginning of the transmission (i.e., the transmitted signal becomes $[s[N_s - L + 1], s[N_s - L + 2], \cdots, s[N_s], \mathbf{s}]$) and if the receiver discards the first $L$ received samples (i.e., cyclic prefix removal), the input-output relation can be expressed as

$$\tilde{\mathbf{y}} = \tilde{\mathbf{H}}_c \mathbf{s} + \mathbf{w} \tag{15}$$

where $\tilde{\mathbf{H}}_c$ becomes a circulant matrix which has a number of important characteristics. Specifically, $\tilde{\mathbf{H}}_c$ can be equivalently expressed as $\tilde{\mathbf{H}}_c = \mathbf{F}^H \mathbf{G} \mathbf{F}$, where $\mathbf{F}$ is a discrete Fourier transform (DFT) matrix of size $N_s$ and $\mathbf{G}$ is a diagonal matrix where its elements are computed from $\tilde{h}_l, \forall l$. Due to this reason, one can precode $\mathbf{s}$ as $\bar{\mathbf{s}} = \mathbf{F}^H \mathbf{s}$ before transmission (i.e., inverse fast Fourier transform (IFFT) operation) and then perform post processing the received signal with $\mathbf{F}$ which consequently yields [17], [18]

$$\tilde{\mathbf{y}} = \mathbf{F}^H \mathbf{G} \mathbf{F} \mathbf{F}^H \mathbf{s} + \mathbf{w} \Rightarrow \mathbf{F} \tilde{\mathbf{y}} = \mathbf{G} \mathbf{s} + \tilde{\mathbf{w}} \Rightarrow \hat{s}[m] = s[m] + \frac{g[m]^* \tilde{w}[m]}{|g[m]|^2}, \quad m = 1, 2, \cdots, N_s \tag{16}$$

where $\tilde{\mathbf{w}} = \mathbf{F} \mathbf{w} = [\tilde{w}[1], \tilde{w}[2], \cdots, \tilde{w}[N_s]]$ and each entry of $\tilde{\mathbf{w}}$ is still an i.i.d zero mean circularly symmetric complex Gaussian (ZMCSCG) random variable with variance $\sigma^2$, $g[m]$ is the $m$th diagonal element of $\mathbf{G}$ (i.e., the channel gain corresponding to the $m$th symbol) and $\hat{s}[m]$ is the estimate of $s[m]$. As we can see $\mathbf{G} \mathbf{s}$ is a parallel channel where each element of $\mathbf{s}$ does not experience any ISI signal. Since $\mathbf{F}$ is a Fourier matrix, one can interpret $\bar{\mathbf{s}}$ as the time domain version of the frequency domain signal $\mathbf{s}$. For this reason, the precoding and post processing operations of the OFDM transmission are usually stated as a signal transformation from frequency domain to time domain and vice verse, respectively.

Now when the receiver has $N$ antennas (SIMO system), each antenna will transform its received signal into a parallel channel as in (16) to coherently combine $\hat{s}[m]$ with the maximum ratio combining (MRC) approach. By doing so, one can express the estimate of $s[m]$ as

$$\hat{s}[m] = s[m] + \frac{\tilde{\mathbf{g}}[m]^H \tilde{\tilde{\mathbf{w}}}[m]}{|\tilde{\mathbf{g}}[m]|^2}, \quad m = 1, 2, \cdots, N_s \tag{17}$$

where $\tilde{\mathbf{g}}[m] = [g_1[m], g_2[m], \cdots, g_N[m]]$ and $\tilde{\tilde{\mathbf{w}}}[m] = [\tilde{w}_1[m], \tilde{w}_2[m], \cdots, \tilde{w}_N[m]]$. As we can see from the above expression, the existing OFDM transmission scheme is able to transmit $N_s$ symbols in $T_b \triangleq (N_s + L)T_s$ seconds. And the SNR of each symbol depends on $N$; increasing (decreasing) $N$ increases (decreases) the SNR of the symbol $s[m]$.

## B. Problem Statement

To better present the objective of the work, we consider the time-domain input-output expression given in (13). When $N \geq L$ and $\mathbf{h}_n = [h_{n0}, h_{n1}, \cdots, h_{n(L-1)}]$ (i.e., the elements of the channel vector defined in (7) corresponding to the $n$th receive antenna) are independent random variables, the received samples obtained at time $LT_s$ in all antennas can be expressed as

$$\mathbf{y}_L = \tilde{\tilde{\mathbf{H}}} \mathbf{s}_{(1:L)} + \mathbf{w}_L \tag{18}$$

where $\tilde{\tilde{\mathbf{H}}} = [\tilde{\mathbf{h}}_1; \tilde{\mathbf{h}}_2; \cdots; \tilde{\mathbf{h}}_N]$, $\mathbf{w}_L = [w_1[L], w_2[L], \cdots, w_N[L]]$ and $\mathbf{s}_{(1:L)} = [s[1], s[2], \cdots, s[L]]^T$ with $\tilde{\mathbf{h}}_n = [\tilde{h}_{n0}, \tilde{h}_{n1}, \tilde{h}_{n2}, \cdots, \tilde{h}_{n(L-2)}, \tilde{h}_{n(L-1)}]$ is the channel vector defined in (12) corresponding to the $n$th receive antenna. Similarly, the received samples at $2LT_s$ can be expressed as

$$\mathbf{y}_{2L} = \tilde{\tilde{\mathbf{H}}} \mathbf{s}_{(L+1:2L)} + \mathbf{w}_{2L}. \tag{19}$$

The expressions (18) and (19) have the same mathematical structure as that of the MIMO channel. Furthermore, one can decode $\mathbf{s}_{(1:L)}$ and $\mathbf{s}_{(L+1:2L)}$ from (18) and (19) efficiently using different existing MIMO detection schemes such as ZF, MRC etc. This demonstrates that the received samples $\mathbf{y}_i, i \neq kL$ are utilized just to improve the SNR of the transmitted symbols $\mathbf{y}_i, i = kL$. In fact, the relation between upper-bound rate, SNR, $T_b$ and $B$ is

$$R = \log_2\left(1 + \frac{L\gamma_0}{\sigma^2}\right) \approx \frac{N_s}{T_b} \frac{1}{B} \log_2\left(1 + \frac{L\gamma_0}{\sigma^2}\right) \tag{20}$$

where $\gamma_0$ is the equivalent SNR of the transmitted symbols $s_i, \forall i$ when they are decoded by employing $\mathbf{y}_i, i = nL$ only. Furthermore, for fixed $T_b$ and $B$, as each sub-carrier transmits only one symbol, the total transmitted symbols in an OFDM frame is $N_s$ (i.e., number of transmitted symbols is the same as that of sub-carriers). Thus, a linear increase in $N_s$ (number of transmitted symbols) increases the rate "linearly" whereas, a linear increase in SNR increases the rate "only logarithmically". Therefore, for a given $T_b$ and $B$, increasing $N_s$ is more advantageous than increasing $L\gamma_0$.

The above discussions show that the existing transmission and decoding approach employs $\mathbf{y}_i, i \neq kL$ just to improve SNR which is not the best strategy from the achievable rate point of view (i.e., the benefits of the time slots $\mathbf{y}_i, i \neq kL$ are not exploited efficiently). This motivates us to set the following objective: For the given symbol constellation, $B$ and $T_b$, is it possible to transmit more than $N_s$ symbols and reliably decode all of the symbols in a symbol-by-symbol basis like in the OFDM approach while achieving better achievable rate than that of the existing OFDM transmission approach? In other words, can we take advantage of the inefficiently used symbol time slots $i \neq kL, \forall k$ to transmit more data symbols while ensuring that the rate is higher than that of the existing OFDM approach. If yes, how can the transmission strategy be designed and to what extent can the rate be increased?

## IV. PROPOSED OFTN TRANSMISSION APPROACH

This section describes the proposed OFTN transmission and data detection approach for SIMO wireless systems to address the above objective. The extension of the proposed design for different system setups will be discussed in the next section.

### A. Transmission Scheme

One can observe from the discussions in the above section that the OFDM transmission scheme transmits only one symbol in $T_s$ seconds. However, as our objective is to transmit more than one data symbol in $T_s$ seconds, we may need to find a way to incorporate additional data symbols per $T_s$. In this regard, we split the overall bandwidth $B$ into $L$ equivalent sub-bands where each of them has bandwidth of $B_L \triangleq \frac{B}{L}$ (i.e., traditional frequency division multiple access where the symbol period of each of the sub-bands is the same as the delay spread of the channel)[6]. With the bandwidth $B_L$, in fact, the existing OFDM transmission approach is able to transmit $N_L \triangleq \frac{N_s}{L}$ data symbols in $T_b$ seconds. Our objective will now be to transmit more than $N_L$ data symbols in the same bandwidth which is detailed as follows.

For better exposition, we assume that the transmitted symbols, $d_0, d_1, d_2, \cdots$ are mapped into period $\tilde{T}_s$, and the transmitter and receiver filters employ bandwidth $B_L = \frac{1}{T_d}$ and have a support duration of $\tilde{Q}$ measured in terms of $\tilde{T}_s$[7]. As in the above section, we start with the SISO system, and the result is then extended to SIMO. With the above settings, we will have the following

---

[6]This can be performed by employing an appropriate scheduling algorithm.

[7]Note that this can be interpreted as oversampling of the transmitted signal $P$ times where, unlike the traditional oversampling with zero padding, the oversampled signals also contain independent information symbols which will be clear in the sequel.

transmitted data signals at time instant $l\tilde{T}_s$

$$x_l \triangleq x(l\tilde{T}_s) = \sum_{k=-\frac{\tilde{Q}}{2}}^{\frac{\tilde{Q}}{2}} \tilde{g}(l\tilde{T}_s - k\tilde{T}_s)d_{l+k}. \tag{21}$$

We would like to emphasize here that the bandwidth of the transmitted signal $x(l\tilde{T}_s)$ is determined by the bandwidth of the filter $\tilde{g}(t)$ which is still $B_L$. For better understanding of the proposed transmission, we assume that $\tilde{g}(t)$ is a rectangular pulse shaping filter of width $T_d$. With $\tilde{g}(t) = \text{rect}(t)$ (i.e., $P = \tilde{Q}$) and after appropriate time shifting, $x(l\tilde{T}_s)$ can be re-expressed as

$$x_l = \sum_{k=l-P+1}^{l} d_k. \tag{22}$$

From this expression, one can understand that the only difference between the existing OFDM transmission scheme and the proposed scheme is that the existing approach sets $d_k = 0, k \neq l$ (i.e., zero padding approach) or $d_k = f(d_{l-P}, d_l, d_{l+P}, \cdots), k \neq l$ (i.e., correlated $d_k$ in the interpolation approach). In the following, we show that it is possible to introduce independent $d_k$ and decode (estimate) them reliably, where the number of introduced independent symbols will depend on $N$ and the number of independent samples of $\mathbf{h}$ which will be clear in the sequel.

The received signal is given by the convolution of the transmitted signal and the channel which can be expressed as

$$r_l = \sum_{m=0}^{P-1} h_m x_{l-m} + \tilde{w}_l \tag{23}$$

where $\tilde{w}_l$ is the additive noise. Once the received signal is passed through the receiver matched filter (i.e., a rect function) and after doing some mathematical manipulations, we will have

$$y_l = \sum_{k=l-P+1}^{l} (r_k + \tilde{w}_k) = \sum_{k=l-P+1}^{l} \sum_{m=0}^{P-1} h_m x_{k-m} + w_l$$
$$= \mathbf{h}^T \mathbf{A} \mathbf{d}_l + w_l = \bar{\mathbf{h}}^T \mathbf{R}^T \mathbf{A} \mathbf{d}_l + w_l \tag{24}$$

where $w_l = \sum_{k=l-\tilde{Q}+1}^{l} \tilde{w}_k$ which is assumed to be a ZMCSCG random variable with variance $\sigma^2$, $\mathbf{d}_l = [d_l, d_{l-1}, \cdots, d_{l-3P+4}, d_{l-3P+3}]^T$, the last equality follows from (7) and $\mathbf{A} = \mathbf{A}_0 \mathbf{A}_1$ with $\mathbf{A}_0 \in \Re^{P \times 2P-1}$ ($\mathbf{A}_1 \in \Re^{2P-1 \times 3P-2}$) is a convolution matrix obtained by utilizing a $P$ row vector of ones (i.e., $\mathbf{A}_0 = \text{convmtx}(\mathbf{1}_P^T, 2P-1)$ and $\mathbf{A}_1 = \text{convmtx}(\mathbf{1}_P^T, 3P-2)$). As we can see from (24), $y_l$ is a linear combination of $3P-2$ symbols (i.e., $d_{l-3\tilde{Q}+3}$ to $d_l$) which has the same dimension as the convolution of the transmit and receive filters, and the channel vector.

*B. Number of Reliably Decoded Symbols*

When the receiver has $N$ antennas, one can stack $y_l$ of all antenna elements to get

$$\mathbf{y}_l = \bar{\mathbf{H}}^T \mathbf{R}^{1/2} \mathbf{A} \mathbf{d}_l + \mathbf{w}_l = \mathbf{Q}^{1/2} \bar{\mathbf{H}}_I \mathbf{R}^{1/2} \mathbf{A} \mathbf{d}_l + \mathbf{w}_l = \bar{\bar{\mathbf{H}}} \mathbf{A} \mathbf{d}_l + \mathbf{w}_l \tag{25}$$

where $\mathbf{Q} \in \mathcal{C}^{N \times N}$ is the spatial channel correlation matrix of all receive antennas, $\bar{\mathbf{H}} = [\bar{\mathbf{h}}_1, \bar{\mathbf{h}}_2, \cdots, \bar{\mathbf{h}}_N]^T = (\mathbf{Q}^{1/2} \bar{\mathbf{H}}_I)^T$, each entry of $\bar{\mathbf{H}}_I$ is assumed to be an i.i.d ZMCSCG random variable with unit variance, $\bar{\bar{\mathbf{H}}} \triangleq \mathbf{Q}^{1/2} \bar{\mathbf{H}}_I \mathbf{R}^{1/2}$ and $\mathbf{w}_l = [w_{1l}, w_{2l}, \cdots, w_{Nl}]$ with $\bar{\mathbf{h}}_n$ as $\bar{\mathbf{h}}$ of (7) corresponding to the $n$th antenna. One can model different propagation characteristics of a wireless environment by setting different values of $\mathbf{Q}$ and $\mathbf{R}$. For instance, a communication environment having few scatterers can be represented by setting a very low rank $\mathbf{R}$. Furthermore, when the antenna spacing is not sufficiently large to experience independent channel coefficients, the spatial correlation matrix $\mathbf{Q}$ will be different from a diagonal matrix. For these reasons, we believe that (25) can be utilized for different wireless environments just by tuning $\mathbf{Q}$ and $\mathbf{R}$.

Since (25) can be considered as a typical MIMO system, the decodability of $\mathbf{d}_l$ with a linear decoder (estimator) and symbol-by-symbol basis entirely depends on the rank of $\bar{\bar{\mathbf{H}}} \mathbf{A}$. And since, $\text{rank}(\bar{\bar{\mathbf{H}}} \mathbf{A}) \leq P$, the number of independent symbols of $\mathbf{d}_l$ cannot be more than $P$. When the maximum number of independent transmitter symbols $(S)$ is less than $P$, one can set the first $S$ coefficients of $\mathbf{d}_l$ to be the desired symbols while leaving the others $0$. By doing so, $\mathbf{A}$ becomes an upper triangular matrix where in such a case $\text{rank}(\bar{\bar{\mathbf{H}}} \mathbf{A}) = \text{rank}(\bar{\bar{\mathbf{H}}}) = \min(\text{rank}(\mathbf{Q}), \text{rank}(\mathbf{R}))$. Therefore, we can express the maximum number of independent transmitted symbols as

$$S = \min(\text{rank}(\mathbf{Q}), \text{rank}(\mathbf{R})) \leq P. \tag{26}$$

From this expression, we can understand that the maximum number of transmitted symbols depends on the ranks of the channel covariance matrices $\mathbf{Q}$ and $\mathbf{R}$. Furthermore, when these two matrices have full ranks (i.e., favorable propagation conditions), we can reliably estimate $P$ independent data symbols of $\mathbf{d}_l$. The advantage of the proposed transmission approach compared to that of the existing one (i.e., OFDM) is that it allows the transmission of $S \leq P$ symbols in $(3P - 2)\tilde{T}_s$ interval whereas, the existing approach transmits only $3$ symbols in the same duration. Thus, in a favorable propagation environment, the proposed transmission scheme is attractive especially when $P$ and $N$ are large.

To show the impact of $N$ on the decodability of $d_{li}$ pictorially, we set $\mathbf{Q} = \mathbf{I}$, $\mathbf{R} = \mathbf{I}$ and considered binary phase shift keying (BPSK) transmitted symbols $d_{li}$. Fig. 1 shows the eye diagram of the estimate of $d_{li}$ obtained by applying ZF receiver for different values of $N$ and $\gamma \to \infty$ (SNR) which is defined as $\gamma = \frac{\mathrm{E}|d_{li}|^2}{\sigma^2}$ while setting $P = 8$. One can notice from this figure that the eye diagram of the estimated data resembles that of the true one (i.e., with values $\pm 1$) when $N = P$. However, when $N < P$, the shape of the eye diagram is not improved even if the SNR goes to infinity. This is due the fact that in such a case the received signal does not have sufficient degrees of freedom (DoF) to reliably detect the transmitted symbols. This implies that the maximum number of transmitted symbols in $T_d$ seconds will be $\min(N, P)$. To verify this, we have also plotted the uncoded bit error rate (BER) of the proposed transmission approach for different values of $N$ as shown in Fig. 2. From this figure, one can also observe that when $N < P$, there is a BER floor and increasing SNR does not help avoid this undesirable effect[8]. However, the BER decreases consistently with SNR when $N \geq P$.

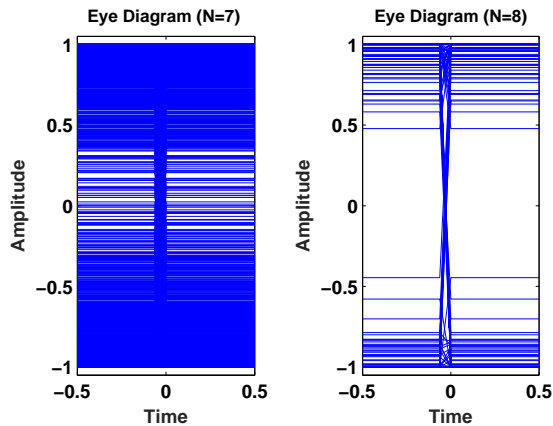 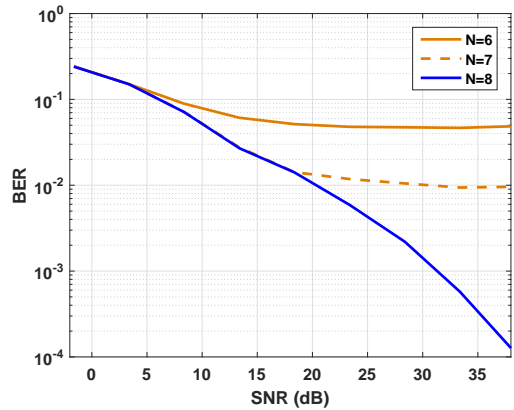

Fig. 1: Eye diagram of proposed design when $P = 8$ and $\gamma \to \infty$.

Fig. 2: The uncoded BER of proposed design for $P = 8$ and $N = [5, 6, 7]$.

## V. DATA DETECTION AND PERFORMANCE ANALYSIS

As can be noticed from the discussions of the previous section, we are able to transmit $S$ symbols in $(3P - 2)\tilde{T}_s$. This section discusses some of the possible data detection techniques to recover $\mathbf{d}_l$ from $\mathbf{y}_l$. In this regard, we discuss three possible approaches, namely MRC, ZF and zero-forcing successive interference cancellation (ZF-SIC) as follows.

---

[8]Note that the eye diagram of Fig. 1 (right side) is not exactly the same as that of the true one. This is because some of the channel realizations of $\bar{\mathbf{H}}$ are not well conditioned which creates some non-zero residual interference on the estimate of $d_{li}$. However, such residual interference will not fundamentally affect the achievable BER (see Fig. 2).

From the existing OFDM transmission, one of the most widely used simple receiver designs is the MRC. For the proposed OFTN transmission, one can design the MRC beamformer as

$$\mathbf{B} = \mathbf{A}^{-1}(\mathbf{R}^{1/2})^{\dagger}\bar{\mathbf{H}}_I^H(\mathbf{Q}^{1/2})^{\dagger}. \tag{27}$$

And for the ZF beamformer, we can design $\mathbf{B}$ as

$$\mathbf{B} = (\mathbf{Q}^{1/2}\bar{\mathbf{H}}_I\mathbf{R}^{1/2}\mathbf{A})^{\dagger}. \tag{28}$$

As can be seen from these two expressions, both the MRC and ZF beamformings have matrix inversion operations. However, for the MRC approach $\mathbf{A}^{-1}$, $\mathbf{R}^{\dagger}$ and $\mathbf{Q}^{\dagger}$ can be computed offline as they are constant design coefficients. For this reason, the MRC beamforming does not need to perform inverse operation when the channel is updated. However, the ZF beamformer indeed requires such an operation whenever the channel is varied (i.e., once per channel coherence time). For this reason, MRC beamforming is computationally less expensive than the ZF counterpart. With these beamforming matrices, we can recover $\mathbf{d}_l$ as

$$\hat{\mathbf{d}}_l = \mathbf{B}\mathbf{y}_l. \tag{29}$$

The main drawback of the MRC and ZF beamforming approaches is both of them require to perform the matrix inverse operation on $\mathbf{A}$ which is an upper triangular matrix. For example, when $S = P = 4$, we will get (30)

$$\mathbf{A}_{(4\times 4)} = \begin{bmatrix} 1 & 2 & 3 & 4 \\ 0 & 1 & 2 & 3 \\ 0 & 0 & 1 & 2 \\ 0 & 0 & 0 & 1 \end{bmatrix}. \tag{30}$$

Consequently, the recovered symbols $\hat{\mathbf{d}}_l$ experience quite different desired signal power which consequently leads to unbalanced SINR and lower total achievable rate. To alleviate this drawback, we propose a combined ZF and SIC (i.e., ZF-SIC) based data detection approach which employs two steps: In the first step, we compute

$$\hat{\mathbf{d}}_l = (\bar{\bar{\mathbf{H}}}^H\bar{\bar{\mathbf{H}}})^{-1}\bar{\bar{\mathbf{H}}}^H\mathbf{y}_l = \mathbf{A}\mathbf{d}_l + \bar{\mathbf{w}}_l \tag{31}$$

where $\bar{\mathbf{w}}_l = (\bar{\bar{\mathbf{H}}}^H\bar{\bar{\mathbf{H}}})^{-1}\bar{\bar{\mathbf{H}}}^H\mathbf{w}_l$. Now since the $S$th element of $\hat{\mathbf{d}}_l$ is interference free, it can be decoded independently. In the second step, we subtract $d_{lS}$ from the $(S-1)$th element of $\hat{\mathbf{d}}_l$ and decode (i.e., $d_{l(S-1)}$). This is because the $(S-1)$th element of $\hat{\mathbf{d}}_l$ contains only one interference term which is known (i.e., $d_{lS}$). Finally, we repeat these two steps until all symbols are decoded.

The remaining question is to assess whether the proposed transmission strategy will always achieve better performance than that of the OFDM or not? Addressing this question for a general $\mathbf{Q}$ and $\mathbf{R}$ appears to be intractable. In the following, we analyze the average performances of the proposed and OFDM transmission schemes for some practically relevant parameter settings.

*Theorem 1*: When $\mathbf{Q} = \mathbf{I}$ and $\mathbf{R} = \mathbf{I}$, the proposed ZF-SIC based approach (assuming no error propagation) and existing OFDM approach can have the following upper-bound average rates

$$\begin{aligned} R^{Ex} &= \log_2(1 + \gamma^{Ex}), \quad \gamma^{Ex} = \frac{LN}{\sigma^2} \\ R^{Pr} &= L\tilde{\alpha} \log_2(1 + \alpha \gamma^{Pr}), \quad \gamma^{Pr} = \frac{N - L + 1}{\sigma^2} \end{aligned} \quad (32)$$

where $R^{Pr}(R^{Ex})$ is the rate achieved by the proposed (existing) approach, $\tilde{\alpha} = \frac{L}{3L-2}$ and $\alpha = \frac{3L-2}{L^2}$. The coefficient $\tilde{\alpha}$ is introduced as the contribution of any $L$ transmitted symbols in $B_L$ bandwidth will span up to $(3L-2)\tilde{T}_s$ durations due to the channel delay, transmit filter and receive filter, and $L$ is introduced to ensure that the proposed OFTN approach will utilize a total bandwidth $B = LB_L$. And $\alpha$ is the power normalization factor which is introduced to maintain that both the OFDM and OFTN approaches use the same average transmit power. Furthermore, if $\frac{N}{\sigma^2} \gg 1$ and $N \gg L$ (i.e., at high SNR and large $N$), we will have $R^{Pr} - R^{Ex} \geq 0$ when

$$N \geq \max\{L, 2^{\frac{1}{L^2-3L+2}[(3L-2)\log_2(L) - L^2 \log_2(\alpha)]} \sigma^2\}. \quad (33)$$

*Proof:* See Appendix A. ■

The proposed design attempts to model the discrete time equivalent impulse response of the channel using (7). In fact, the number of multipath channel coefficients definitely increases with bandwidth. For instance, in the ultra wideband communication, one can have $L$ in the order of $200 - 500$, where the channel coefficients often tend to be sparse [17]. The considered model captures the channel characteristics of such an environment by selecting $\mathbf{Q}$ and $\mathbf{R}$ appropriately. This justifies the need to examine the performance of the proposed design for arbitrary correlation matrices. This analysis is not trivial mainly because the SNR expression incorporates an inverse matrix containing random variables which is difficult to handle. However, when $N$ is large, by utilizing the law of large numbers, most of the random components become deterministic which enables us to study the system performance for large $N$ in the following lemma.

*Lemma 1*: When $N \gg L$ (i.e., for a large antenna array), the existing and proposed approaches achieve the following upper-bound average rates

$$R^{Ex} = \frac{1}{N_s} \sum_{i=1}^{N_s} \log_2(1 + \gamma_i^{Ex}), \quad \gamma_i^{Ex} = \frac{N}{\sigma^2} \mathbf{f}_i^T \mathbf{R} \mathbf{R}^T \mathbf{f}_i^*$$

$$R_{ZF}^{Pr} \approx \frac{1}{\tilde{\alpha}} \frac{1}{S} \sum_{i=1}^{S} \log_2(1 + \tilde{\alpha} \gamma_i^{ZF}), \quad \gamma_i^{ZF} = \frac{N}{\sigma^2 \mathbf{E}_{(i,i)}}$$

$$R_{ZF-SIC}^{Pr} \approx \frac{1}{\tilde{\alpha}} \frac{1}{S} \sum_{i=1}^{S} \log_2(1 + \tilde{\alpha} \gamma_i^{ZF-SIC}), \quad \gamma_i^{ZF-SIC} = \frac{N}{\sigma^2 \mathbf{C}_{(i,i)}} \quad (34)$$

where $R^{Ex}$, $R_{ZF}^{Pr}$ and $R_{ZF-SIC}^{Pr}$ are the rates achieved by the existing, proposed ZF and proposed ZF-SIC approaches, respectively, $\tilde{\alpha} = \frac{3L-2}{LS}$, $\mathbf{C} = (\mathbf{R}^H \mathbf{R})^{-1}$ and $\mathbf{E} = (\mathbf{A}^H \mathbf{R}^H \mathbf{R} \mathbf{A})^{-1}$. As expected, when $\mathbf{Q} = \mathbf{I}$, $\mathbf{R} = \mathbf{I}$ and $N \gg L$, the rate expressions of (34) is almost the same as that of (32) since $N \approx N - L$.

*Proof:* See Appendix B. ∎

From the results of *Theorem 1* and *Lemma 1*, the following points can be highlighted:

1) In (32), as $\gamma^{Ex} \approx L \gamma^{Pr}$ for $L \geq 3$ and medium to large $N$, one can interpret this equation that the existing OFDM approach attempts to ensure that each transmitted symbol has a maximum SNR (i.e., diversity gain achieving strategy). However, the proposed approach increases the number of transmitted symbols while reducing the SNR achieved by each symbol (i.e., multiplexing gain achieving strategy). Nevertheless, with the proposed approach, it is always possible to decrease $S$ while increasing the SNR, for instance by reducing $S$ to half, the SNR can be doubled. To show this fact, we have plotted the achievable $\gamma_{Eqt} = \alpha \gamma^{Pr}$ for different values of $S$ and $N$ in Fig. 3. As can be seen from this figure, the SNR of each symbol can be increased just by decreasing $S$. This demonstrates that there exists a multipath diversity-multiplexing tradeoff for SIMO systems. Having said this, however, exploiting this tradeoff for general coding schemes (for example with Alamouti space time block coding (STBC) [36]) is an interesting open problem which is left for future research.

   We would like to emphasize here that the existence of a diversity-multiplexing tradeoff has been known in the existing transmission scheme when both the transmitter and receiver are equipped with multiple antennas (i.e., MIMO) [16]. However, to the best of our knowledge, we are not aware of any existing work demonstrating the diversity-multiplexing tradeoff which is exhibited by exploiting multipath components for a SIMO system.

2) From (33), one can notice that the proposed approach achieves better performance at large $N$. Furthermore, for fixed $L$ and $\sigma^2$, increasing $N$ increases the gap between the proposed and existing designs. This shows that the proposed design is practically useful especially for massive MIMO systems. Nevertheless, the proposed approach can still be customized for small $N$ by optimizing $S$. However, optimizing $S$ for a given $N$ (possible small) is still an open research topic and is left for future research direction.

3) As the proposed algorithm does not require IFFT operation at the transmitter, it does not suffer from high PAPR which merits the benefit of proposed design compared to OFDM. Furthermore, the main computational load of the proposed algorithm is the matrix inversion operation of (28) which is comparable with the OFDM transmission.

4) The noise samples obtained from (25) are i.i.d which consequently avoids the need of pre-whitened matched filter which has been employed in the existing FTN transmission schemes [4], [37]. The disadvantage of the pre-whitening operation is that it often amplifies adjacent channel interference signals which consequently worsens the achievable BER of the FTN transmission. Furthermore, the pre-whitening phase enables each data symbol to experience ISI symbols which consequently requires complicated decoding operation especially when the number of ISI symbols is large which is not desirable in practice [4].

5) One can also observe from (25) that the desired information is obtained every $3P - 2$ sampling periods. This behavior will also allow to use low cost analog to digital converter (ADC) which is one of the most expensive parts of a receiver unit especially in massive MIMO regime [38]–[40].

Although the proposed ZF-SIC based data detection approach maintains the desired symbols of $\mathbf{d}_l$ to have balanced signal power, the latter approach may suffer from error propagation. In other words, if $d_{lS}$ is decoded incorrectly, it will affect the SNR of the subsequently decoded symbols. One way of addressing this issue is by enabling appropriate precoding operation at the transmitter. In fact $\mathbf{A}$ of (31) is known a priori to the transmitter and it is an upper triangular matrix which is invertible. Due to this reason, it is possible to employ ZF precoding (ZF-P) while ensuring $\mathbf{Ad}_l$ becomes a diagonal matrix. Since such an approach ensures ISI free received signal, the receiver can decode the transmitted symbols in symbol by symbol basis. From this discussion, one can understand that the performance of the ZF-SIC (without error propagation) can be achieved by employing a ZF-P operation at the transmitter which helps reduce the complexity of receiver.

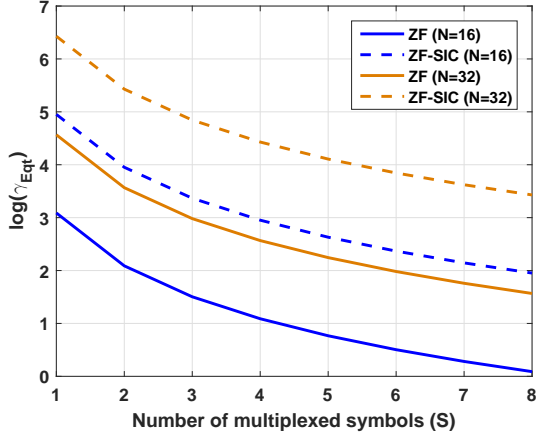 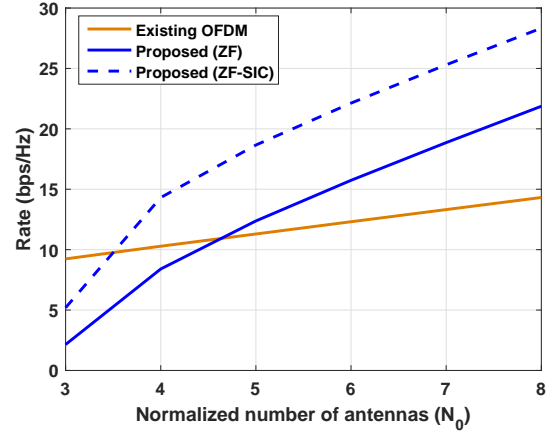

Fig. 3: Logarithm of SNR $\log_2(\gamma_{Eqt})$ when $L = 8$ and $\gamma_{Eqt} = \alpha \gamma^{Pr}$ with $\sigma^2 = 10$ dB.

Fig. 4: Rate of proposed and existing designs when $L = 8$ and $\gamma = 5.3$ dB.

**Complexity**: One can understand from the result of this section that the main complexity of the OFDM approach arises due to the fast Fourier transform (FFT) operation which has complexity $N_s \log(N_s)$. On the other hand, for the proposed OFTN, the main complexity arises to compute $(\bar{\bar{\mathbf{H}}}^H \bar{\bar{\mathbf{H}}})^{-1}$ in (29) of the revised manuscript which has a complexity $O(L^{2.373})$ [41]. In practice since $N_s$ is selected to be much greater than $L$, the complexity of OFDM and proposed OFTN approaches are comparable.

Note that for a fixed bandwidth, one may think of increasing the number of sub-carriers by decreasing the sub-carrier spacing $\Delta f$. In an OFDM system, the symbols which are sent on carrier frequencies with frequency spacing $\Delta f$, the time-frequency product $\Delta t \Delta f$ may need to be constant to achieve interference free sub-channels where $\Delta t$ is the OFDM duration[9]. When we employ the Nyquist transmission, the time-bandwidth product is roughly one [42]. Now if we would like to decrease $\Delta f$, the OFDM transmission duration $\Delta t$ may need to be increased to maintain $\Delta t \Delta f \approx 1$. Thus, for the fixed OFDM duration $\Delta t$, it is practically not possible to decrease the sub-carrier spacing beyond some value.

## VI. ADAPTIVE POWER ALLOCATION WITH EXACT AVERAGE RATES

Up to now we consider the upper-bound rate expressions obtained by existing OFDM and proposed OFTN approaches by assuming equal power allocation policy. In fact, for the given $L$, one can apply the law of large numbers to show that $|\mathbf{h}_{oi}|^2$ and $\bar{\mathbf{H}}_I^H \bar{\mathbf{H}}_I$ are almost constant when $N \gg L$ (i.e., massive MIMO setup) [34]. In such a case, equal power allocation yields close to optimal rate (i.e., power adaptation results in negligible average rate gain). For a small to medium $N$, however, appropriate power adaptation may help improve the average rates of both the OFDM and OFTN approaches since the effect of fading is not negligible. Here we consider the optimal power adaptation strategy which employs the exact average rates as discussed in [43]. For a fading channel with Gaussian transmitted signal and SISO system, the optimal power allocation that maximizes the average rate is formulated as

$$R_{av} = \max \int \log_2\left(1 + \frac{\Gamma(\psi)}{P_{av}}\psi\right)\rho(\psi)d\gamma$$
$$\text{s.t} \int \Gamma(\psi)\rho(\psi)d\psi \leq P_{av} \qquad (35)$$

where $P_{av}$ is the average power budget, $\psi$ is the received signal SNR, $\Gamma(\psi)$ is the allocated power, and $\rho(\psi)$ is the probability density function (PDF) of $\psi$. The optimal power allocation $\Gamma(\psi)$ that maximizes $R_{av}$ satisfies (see (4)-(6) of [43])

$$\frac{\Gamma(\psi)}{P_{av}} = \begin{cases} \frac{1}{\psi_0} - \frac{1}{\psi}, & \psi \geq \psi_0 \\ 0, & \psi < \psi_0 \end{cases} \qquad (36)$$

where $\psi_0$ is the cut-off SNR which is selected to ensure [44]

$$\int_{\psi_0}^{\infty}\left(\frac{1}{\psi_0} - \frac{1}{\psi}\right)\rho(\psi)d\psi = 1. \qquad (37)$$

---
[9]It is also possible to use the term time-bandwidth instead of time-frequency.

Once $\psi_0$ is computed, the optimal average rate is given as

$$R_{av} = \int_{\psi_0}^{\infty} \log_2\left(\frac{\psi}{\psi_0}\right)\rho(\psi)d\psi. \qquad (38)$$

One can observe from (38) that the optimal average rate depends on $\rho(\psi)$[10].

Now if we adopt the steps (35) - (38) to compute the average rates of the OFDM ($R_{av}^{Ex}$) and proposed OFTN ($R_{av}^{Pr}$) transmissions, we will have

$$R_{av}^{Ex} = R_{av}|_{\rho(\psi)=\rho(\psi)^{Ex}} \qquad (39)$$

---

[10]Note that to realize the power adaptation policy, the instantaneous received signal SNR $\psi$ may need to be available at the transmitter which can be obtained from the feedback SNR in frequency division duplex (FDD) system or from the estimated CSI in time division duplex (TDD) system.

where $\rho(\psi)^{Ex}$ is a Gamma distribution with parameters $k = NL$ and $\theta = \frac{1}{\sigma^2}$. This is due to the fact that $2\sigma^2 \gamma^{Ex}$ in Theorem 1 is a Chi-square ($\chi^2$) distribution with $2NL$ degrees of freedom. Furthermore, for the proposed OFTN approach with ZF-SIC (no error propagation), we will have

$$R_{av}^{Pr} = \frac{1}{\alpha} R_{av}|_{\rho(\psi) = \rho(\psi)^{Pr}} \tag{40}$$

where $\rho(\psi)^{Pr}$ is a Gamma distribution with parameters $k = \frac{N-L+1}{2}$ and $\theta = \frac{2\alpha}{\sigma^2}$ when $S = L$.

## VII. OFTN Vs FTN, and Effect of Pulse Shaping Filter

### A. OFTN Vs FTN for SISO Systems

The current paper considers a FTN transmission scheme for wideband channels having multipath components. This will raise a question what is the key difference between the proposed and existing FTN transmission approaches just for SISO additive white Gaussian noise (AWGN) channel which is the focus of many existing works [4]. In fact, at the transmitter side, one can still use the methodology used in Section V to realize FTN transmission (for example with precoding). Furthermore, by stacking the first $\tilde{Q} = L$ received samples taken every $\tilde{T}_s$ and after filtering and doing some manipulations, we can express the filtered signal as

$$\mathbf{y}_l = \mathbf{d}_l + \mathbf{A}_0 \tilde{\tilde{\mathbf{w}}}_l \tag{41}$$

where $\tilde{\tilde{\mathbf{w}}}_l$ noise vectors where each of its entry is i.i.d ZMCSCG random variable each with variance $\sigma^2$ and $\mathbf{A}_0$ is as defined in (24). As we can see (41) contains correlated noise vectors which will share the same drawbacks as that of the existing FTN transmission schemes. This shows that in the case of AWGN or flat fading channels, the proposed design does not have any advantages compared to the existing FTN transmission. This also confirms that the gain achieved by the proposed transmission does not come just because of rectangular pulse shaping filter[11].

We would like to emphasize here that the current paper focuses on a typical OFTN transmission and the corresponding data estimation technique for a wideband SIMO system. The technique used in this paper, however, may not be necessarily optimal. Nevertheless, as will be demonstrated in the simulation section, the proposed technique achieves significantly better transmission rate than that of the existing OFDM technique.

---

[11]We would like to point out here that one can also enable the benefits of the other sampling time received signals of the SIMO channel. However, such an approach will create correlated noise and increase the requirement of ADC speed which will introduce additional complexity to the receiver. Nevertheless, this approach will definitely help to improve performance, and analyzing such a case is beyond the scope of the paper is left for future.

## B. Effect of Pulse Shaping Filter

As can be seen from the discussions in section IV, the proposed approach employs a rectangular pulse shaping filter $g(.)$. When $g(t)$ is different from $\text{rect}(x)$, (22) will be slightly modified which consequently yields different $y_l$. In such a case, however, it is difficult to express $y_l$ in closed form. Nevertheless, as the channel coefficients $h_1, h_2, \cdots h_P$ are not related to $g(t)$, we believe that employing different $g(t)$ would change the gain achieved by the proposed transmission scheme only slightly. Furthermore, one can still employ rect $g(t)$ just by utilizing appropriate guard band like in the OFDM technique.

## VIII. SIMULATION RESULTS

This section provides simulation results, we set $B = 1.25$ MHz, $S = L = P = 8$, $T_d = \frac{L}{B}$ and the carrier frequency is set to 2 GHz. For such settings, we compare the rates obtained by the proposed and existing OFDM transmission approaches. For the OFDM approach, the number of sub-carriers (i.e., FFT size) is set to 128 as in LTE system which is used in 1.4 MHz channel bandwidth transmission [27], [28]. All results are obtained by averaging 10000 channel realizations, the SNR is defined as $\gamma = \frac{\mathrm{E}\{|d_i|^2\}}{\sigma^2}$, and $N = 2^{N_0}$ where $N_0$ is an integer which is chosen to attain the desired $N$.

### A. Comparison of Existing OFDM and Proposed Approaches

This subsection compares the upper-bound rates achieved by the existing OFDM and proposed approaches. Towards this end, we assume that $\mathbf{Q} = \mathbf{I}$ and $\mathbf{R} = \mathbf{I}$. Fig. 4 shows the rates achieved by the proposed and existing designs for different values of $N$ when $\gamma = 5.3$ dB. As we can see from this figure, the proposed design with ZF-SIC detection approach achieves better performance than that of the existing OFDM approach approximately when $N \geq 12$, and the rate gap between the proposed and existing approach increases as $N$ increases[12]. Furthermore, the ZF-SIC data detection approach yields better performance than that of the ZF data detection which is expected.

In the next simulation, we examine the effect of $\gamma$ on the performances of the proposed and existing OFDM transmissions as shown in Fig. 5. One can observe from this figure that the proposed ZF-SIC approach achieves better performance than that of the existing approach when $N \geq 8$ at $\gamma = 15.3$ dB. Furthermore, the required $N$ such that the proposed approaches (ZF and ZF-SIC) achieve better performance than that of the existing approach increases as the SNR decreases. This demonstrates that the proposed approach can still be applied for existing LTE systems where each BS is possibly equipped with few antennas in the high SNR regime.

---
[12]Recall that the performance of ZF-SIC is the same as that of ZF-P.

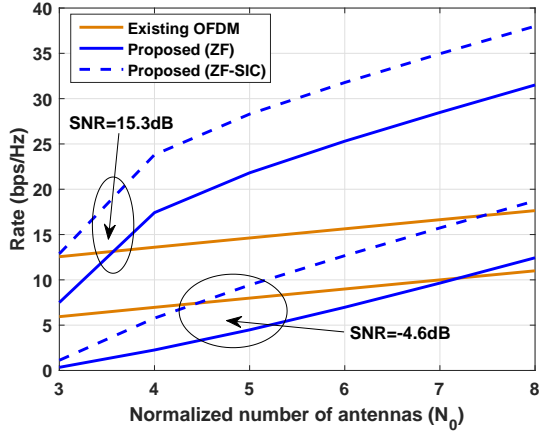 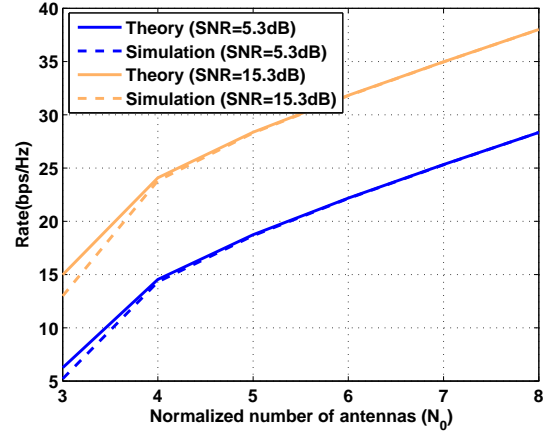

Fig. 5: Effect of SNR on the performances of the Existing OFDM and proposed approaches when $L = 8$.

Fig. 6: Comparison of Theoretical and Simulation rates of the proposed design with ZF-SIC detection approach.

### B. Validation of Theoretical Results

This subsection validates the theoretical results derived in Section V. In this aspect, we demonstrate the theoretical upper-bound rate derived in Theorem 1 for different $N_0$ while setting the SNR as $\gamma = \{5.3, 15.3\}$ dB in Fig. 6. As can be seen from this figure, the results given in Theorem 1 fit well with the simulation results for all antenna array sizes. In Theorem 1, it is stated that the proposed design will achieve better data rate than OFDM transmission when the number of antennas is selected as in (33) which is demonstrated in Fig. 7. For the settings of this figure, one can obtain $N$ of (33) as $N = [16, 8, 8]$ for $\gamma = [-4.6, 5.3, 15.3]$ dB. As can be seen from this figure, the required number of antennas to achieve better performance decreases as SNR increases which demonstrates the benefit of the proposed design even for small antenna array sizes (i.e., $N = P = 8$) at the high SNR regime. Furthermore, the approximated expression of (33) can be regarded as a good estimation for practical system design.

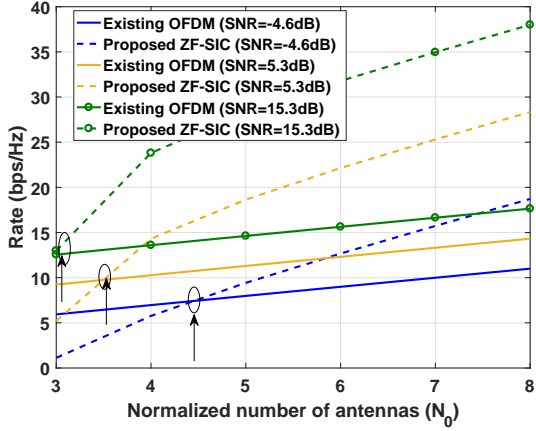 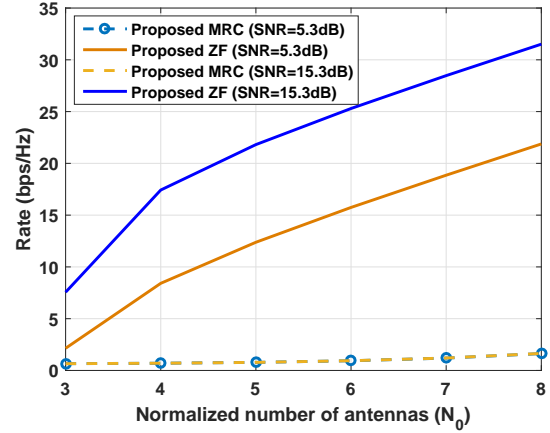

Fig. 7: Validation of the required number of antennas expression of (33).

Fig. 8: Comparison of MRC and ZF beamformings of the proposed design.

## C. Effect of Beamforming Scheme

One can notice from the above results that the ZF-SIC beamformer achieves better performance than that of ZF. However, when $N$ and $L$ are large, one may think of employing a simple MRC beamforming approach to reduce complexity. Thus, it is desirable to compare the performance of ZF and MRC beamforming approaches. In this regard, we have used $\mathbf{R} = \mathbf{I}, \mathbf{Q} = \mathbf{I}$ and $\gamma = \{5.3, 15.3\}$ dB. One can observe from Fig. 8 that ZF beamformer achieves significantly better performance than that of the MRC. Furthermore, increasing the SNR level of the transmission does not help improve the upper-bound rate of MRC. The potential reason for this is that the MRC beamformer does not yield a strictly diagonal $\bar{\mathbf{H}}_I^H \bar{\mathbf{H}}_I$ which will ultimately create non-vanishing ISI terms which are dominant at the very high SNR regime. Consequently, increasing the SNR does not provide meaningful rate improvement.

## D. Desired Rate Regions

One can observe from Fig. 4 - Fig. 7 that the proposed OFTN approach achieves better performance in the region where the average rate is higher than a certain threshold. This raises a concern whether the performance improvement is in a practically relevant rate region or not. In general, it is not trivial to provide exact answer for this concern for a general setup. Here, we try to provide some insight for the setup $\mathbf{Q} = \mathbf{I}$, $\mathbf{R} = \mathbf{I}$ and $L = P = \tilde{Q} = S$. For this setup, one can compare the rates of OFDM and OFTN approaches for fixed $N$ or SNR. In fact, the rate gap between OFTN and OFDM approaches increases with $N$ for fixed SNR. Furthermore, as the proposed OFTN transmission packs $L$ symbols in $L\tilde{T}_s$ seconds (i.e., one symbol period with Nyquist transmission)[13], the actual constellation of these $L$ symbols is dictated by the quantity $\frac{1}{L}R^{Pr}$. These observations motivate us to compare $\frac{1}{L}R^{Pr}$ and $\frac{1}{L}R^{Ex}$ (termed as normalized rate) for different SNR and $L$ by setting the minimum $N$ (i.e., $N = L$) as shown in Fig. 9. One can observe from this figure that the proposed approach achieves better performance for the region $R \geq R_{th}$ and most importantly $R_{th}$ decreases with $L$. For instance, when $L = 12$, $R_{th} \approx 1$ (i.e., BPSK transmission) and when $L = 4$, $R_{th} \approx 3$ (i.e., 8 PSK transmission)[14]. This phenomena renders the proposed OFTN approach advantageous compared to OFDM as $L$ is related to the transmission bandwidth. And for fixed SNR and $N$, it is always possible to choose $L$ optimally while ensuring that the OFTN approach reaps the maximum benefit compared to OFDM. One simple, albeit suboptimal, strategy would be to choose $L$ as the maximum value that ensures the OFTN approach to have a normalized rate at least 1 (i.e., BPSK constellation) which can be computed by gradually decreasing $L$ starting from $L = N$[15].

From this discussion, one can understand that the proposed OFTN approach is flexible and can be applied for high and low data rate applications.

### E. Tightness of the Upper-bound Rates and Effect of Power Adaptation

In this simulation, we examine the tightness of the rate expressions $R^{Ex}$ and $R^{Pr}$, and the effect of power adaptation on the performances of OFDM and proposed OFTN approaches. As stated in Section VI, power adaptation may help improve the average rates of both the OFDM and OFTN approaches for a small to medium $N$ since the effect of fading is not negligible. For this reason, we employ $L = 8$, $\mathbf{R} = \mathbf{I}, \mathbf{Q} = \mathbf{I}$, $\gamma = 5.3$ dB and small to medium $N$, and compare $\{R_{av}^{Ex}, R_{av}^{Pr}\}$ (i.e., optimal power adaptation), and $\{R^{Ex}, R^{Pr}\}$ (i.e., equal power allocation) as shown in Fig. 10 where we compute the integral terms (37) and (38) numerically. One can observe from this figure that the rate gap between the proposed approach and that of OFDM increases as $N$ increases for fixed $L$. Furthermore, this gap is almost similar with and without power adaptation. In addition, the upper-bound rate expressions presented in Theorem 1 is tight and power adaptation provides almost no gain both for OFDM and proposed OFTN approaches since $R_{av}^{Ex} \approx R^{Ex}$ and $R_{av}^{Pr} \approx R^{Pr}$.

---

[13]Note that the ratio $\frac{L}{3L-2}$ is just due to the rate inefficiency because of padding $2L - 2$ zero samples every $3L - 2$ sampling periods.

[14]Note that the region with a normalized rate less than one is not desirable as it does not represent a realistic signal constellation.

[15]Recall that a slightly better performance optimization can be achieved by further optimizing $S$ for fixed $L$.

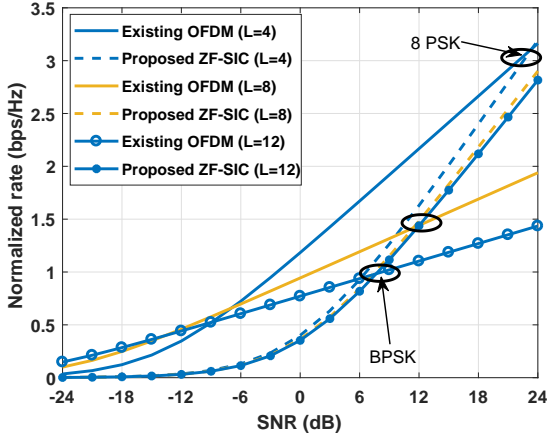
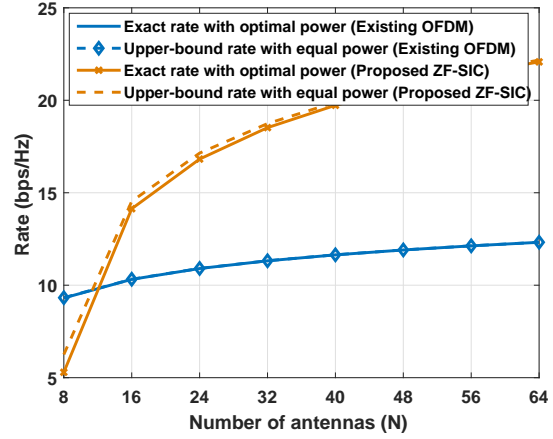

Fig. 9: Comparison of the normalized rates of OFTN and OFDM approaches for different SNR and $L$.

Fig. 10: Average rates achieved by OFDM and OFTN approaches with optimal and equal power allocations when $\gamma = 5.3$ dB.

## F. Effect of Channel Correlations

This simulation examines the effect of temporal and spatial channel correlations on the performances of the proposed design. For the spatial channel correlation matrix, we employ the well known exponential correlation model as $\mathbf{Q}_{i,j} = \rho^{|i-j|}$, where $0 \leq \rho < 1$. The exponential model is physically reasonable in a way that the correlation between two transmit antennas decreases as the distance between them increases [45]. Furthermore, this model is a widely used antenna correlation model for urban area communications where traffic is usually congested [46]. Fig. 11 shows the performances of the proposed designs for $\rho = \{0.1, 0.5, 0.95\}$ with $\gamma = 5.3$ dB and $\mathbf{R} = \mathbf{I}$. One can observe from this figure that the performance of the proposed design degrades when $\rho$ increases for all $N$. However, for fixed $\rho$, increasing $N$ not only helps increase the upper-bound rate but it also help decrease the rate gap between the correlated channel and the i.i.d (i.e., $\rho = 0$). The main reason for this observation is that for fixed $\rho$, increasing $N$ will increase the number of eigenvalues of $\mathbf{Q}$. And for this simulation since $S = P$, the minimum normalized power gain is the same as the $P$th maximum normalized eigenvalues of $\mathbf{Q}$ which increases with $N$. To show this, we plot the normalized eigenvalues of $\mathbf{Q}$ for $\rho = 0.5$ and different values of $N$ in Fig. 12. As can be seen from this figure $\frac{\lambda_P}{\lambda_1}$ increases as $N$ increases. This also validates the result of Lemma 1 which states that the performance of the proposed design does not depend on $\mathbf{Q}$ when $N \gg L$ and $\text{rank}(\mathbf{Q}) \approx P$.

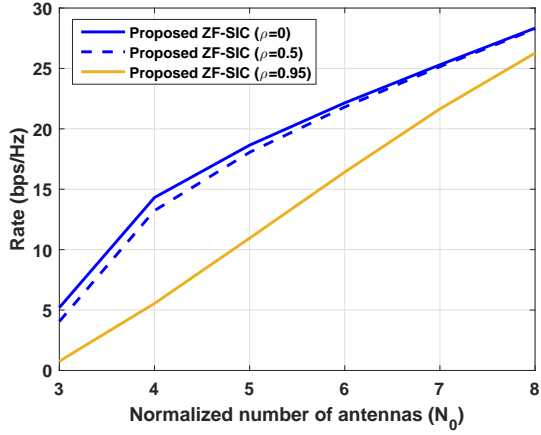 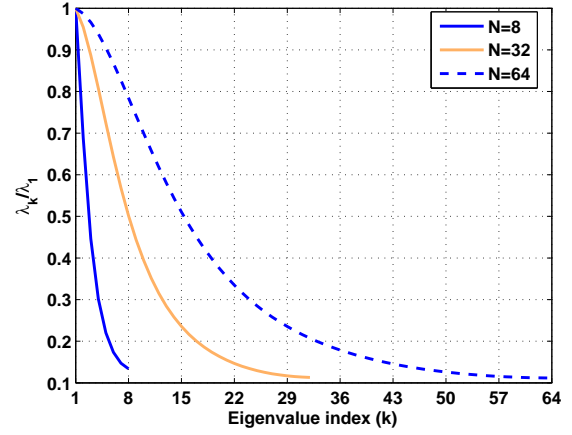

Fig. 11: Performance of the proposed ZF-SIC detection for different $\rho$.

Fig. 12: Normalized Eigenvalue distribution $\frac{\lambda_i}{\lambda_1}$ (decreasing order) when $\rho = 0.5$.

A number of approaches have been proposed to model the temporal channel correlation matrix $\mathbf{R}$. One most commonly employed model is the COST 207 for urban environment where $\mathbf{R}$ is a diagonal matrix with its elements decay exponentially [47], [48]. Under this model, $\mathbf{R}_{i,i}$ can be chosen as (see (5), (7) and (19) of [47])

$$\mathbf{R}_{i,i} = \kappa \int_{\tau_{(i-1)}}^{\tau_i} e^{-Bx} dx \tag{42}$$

where $\kappa$ is a normalization factor which depends on the carrier frequency $f_c$ and is chosen to ensure $\mathbf{R}_{0,0} = 1$ and $B$ is the transmission bandwidth. With this model, the diagonal elements of $\mathbf{R}$ becomes $[1, 0.3679, 0.1353, 0.0498, 0.0183, 0.0067, 0.0025, 0.0009]$. As we can see, the diagonal elements of $\mathbf{R}$ decrease with $i$. This will ultimately decrease the effective rank of $\mathbf{R}$. For this reason, one can expect reduced data rate when $\mathbf{R}$ is designed as in (42) compared to that of $\mathbf{R} = \mathbf{I}$. The simulation results demonstrating this fact has been omitted for conciseness.

From the result of this section, we can understand that the performance of the proposed OFTN transmission approach varies from one scenario to another. Furthermore, one can improve the performance of the proposed design by selecting $S$ adaptively based on the structures of $\mathbf{R}$ and $\mathbf{Q}$. Nevertheless, how to design optimal $S$ for different scenario is left for future research

## IX. OPEN RESEARCH TOPICS

The paper focuses on the OFTN transmission for SIMO system by assuming typical multipath channel models. In this section, we first discuss some of the open research topics for the SIMO system. Then, we discuss how one can utilize the proposed OFTN transmission for other single user and multiuser systems, and the associated research problems.

### A. Single User SIMO Systems

In a typical OFDM system, the achievable rate can be computed by assuming equal power allocation, or by optimizing the powers allocated to all sub-carriers. In fact, applying a suitable power allocation strategy will definitely help improve the achievable rate. However, this improvement is achieved at the expense of additional computational cost. In the current paper, we compare the performance of the proposed transmission with that of the OFDM with equal power allocation just for simplicity. The comparison of the OFDM transmission and the proposed approach under optimized power allocation is an interesting open research problem.

In general, the covariance matrices $\mathbf{Q}$ and $\mathbf{R}$ of (25) depend on different factors including the deployment environment, carrier frequency and number of scatterers [49]. Furthermore, there are different multipath channel covariance matrix models for practically relevant wireless environments like Extended Pedestrian A (EPA), Extended Vehicular A (EVA), or Extended Typical Urban (ETU). In the current paper, however, we have provided numerical results for the most commonly adopted channel covariance matrices in the literature such as $\mathbf{R} = \mathbf{I}$ (i.e., multipath coefficients have equal average power), to simplify the analysis and to get an insight about the performance of the proposed approach. Furthermore, the current paper assumes that the channel state information (CSI) is known to the receiver perfectly. In practice, however, the CSI may need to be estimated, and often known imperfectly. Thus, the performance comparison of the proposed approach and OFDM with perfect and imperfect channel model parameters of the above environments is still an open problem.

The current paper assumes that the receive antennas are collocated (i.e., single cell setup). However, it may be the case that the antennas may be placed at different locations like in the distributed (network) MIMO system [50]. In general, the proposed approach can still be applied for the latter system. And in such a system, the elements of $\mathbf{R}$ and $\mathbf{Q}$ will be different from that of the collocated case. Nevertheless, comparing the proposed approach with that of OFDM for network MIMO system case is an interesting questions which needs further research.

*B. Single User MISO and MIMO Systems*

In this subsection, we discuss the potential extensions and open research issues of the proposed data transmission and detection approach for MISO and MIMO channels. In fact, if the transmitter is equipped with multiple antennas, multiplexing gain can be achieved. In this regard, one may think of two possibilities: the first possibility is to simply transmit independent information from each antenna. The second possibility is to utilize appropriate precoding operation. The former possibility can be realized like in the aforementioned sections which consequently help scale the data rate with the number of transmitter antennas $M$ in a favorable propagation condition[16]. However, the receiver requires at least $N = MS$ antennas to reliably estimate the transmitted signal with linear receivers. Although, the latter possibility is well understood from existing communication system, the beamforming design with the proposed OFTN transmission is still not understood and it is left for future work.

The remaining issue is whether the proposed approach can be extended for the MISO channel? As stated previously, when the transmitter has multiple antennas, each antenna can transmit independent information. However, for each transmit antenna, $S$ receive antennas are required. This leads to a fundamental question on how to realize the proposed OFTN transmission and detection (estimation) approach for MISO systems. From the existing OFDM transmission approach, it is well known that the rates achieved by interchanging the roles of the transmitter and receiver are the same (i.e., there is a so called uplink-downlink rate duality (see for example [51]–[53])). However, showing the rate duality of MISO and SIMO channels under the proposed OFTN transmission and data detection is still not understood well and it is left for future research.

*C. Multiuser Systems*

The proposed data transmission and detection approach is designed for a single user multi-antenna system with multipath components. In fact, multipath components exist almost in all wireless wideband channels which justifies that the proposed transmission approach can potentially be applied for different system setups. For example, it is applicable to a wideband multiuser (massive) MIMO system where the BS is equipped with several antennas whereas, each UE is equipped with single antennas. For such a system, each UE can apply the proposed transmission approach. Consequently, the rate achieved by each UE increases with $P$. Nevertheless, examining the performance of the proposed design for different system setups is one of the interesting research directions. The performance can be studied by taking into account the effects of pulse shaping filter $g(t)$ and error propagation in ZF-SIC receivers.

---

[16]Here a favorable propagation condition means that the spatial channel covariance matrix between any transmitter and all receiver antennas is close to $\mathbf{I}$ (or well conditioned).

## X. CONCLUSIONS

In this paper, we propose a novel and simple OFTN transmission and detection approach for a wideband SIMO system with $L$ multipath components. The proposed OFTN transmission exploits the multiplexing gain obtained by the inherent characteristics of multipath components of wideband channels. By doing so, the proposed design achieves significantly higher transmission rate than the conventional Nyquist transmission approach (i.e., OFDM). The superiority of the proposed approach has been shown theoretically and demonstrated through extensive computer simulations. It is also shown that the capacity gap between the proposed approach and that of the OFDM increases as the number of receiver antennas increases for fixed $L$ which implies that the new approach may have practical significance for large antenna array receiver systems. The proposed method can be extended to (massive) MIMO systems. Having said this, however, extending the proposed approach for a MISO system while achieving the same capacity as that of the SIMO system is a non trivial fundamental challenge.

## APPENDIX A

### PROOF OF *Theorem 1*

For the existing OFDM transmission, by utilizing (7) and (11), the channel gain of the $s$th sub-carrier of each receive antenna, $g_s$ of (16) can be expressed as

$$g_s = \mathbf{f}_s^H \mathbf{R}^{1/2} \bar{\mathbf{h}} = \bar{\mathbf{h}}^T (\mathbf{R}^{1/2})^T \mathbf{f}_s^* \tag{43}$$

where $\mathbf{f}_s$ is an $L$ sized truncated DFT matrix with overall size $N_s$ [18], [29]. Using (25) and after some manipulations, $\mathrm{E}\{|\tilde{\mathbf{g}}_s|^2\}$ becomes

$$\mathrm{E}\{|\tilde{\mathbf{g}}_s|^2\} = \mathrm{E}\{|\mathbf{Q}^{1/2}\bar{\mathbf{H}}_I(\mathbf{R}^{1/2})^T\mathbf{f}_s^*|^2\} = N\mathbf{f}_s^T\mathbf{R}\mathbf{f}_s^*, \quad \Rightarrow \gamma_{Ex} = \frac{\mathrm{E}\{|\tilde{\mathbf{g}}_s|^2\}}{\sigma^2} = \frac{N}{\sigma^2}\mathbf{f}_s^T\mathbf{R}\mathbf{f}_s^* \tag{44}$$

where the second equality uses $\text{tr}\{\mathbf{Q}\} = N$ for any $\mathbf{Q}$. When $\mathbf{R} = \mathbf{I}$, we will get $\gamma_{Ex}$ of (32).

For the proposed approach, when the ZF beamforming with SIC is used, we will have

$$\gamma_i = \frac{1}{\sigma^2 \mathrm{E}\{([\bar{\mathbf{H}}_I^H \bar{\mathbf{H}}_I]^{-1})_{i,i}\}}. \tag{45}$$

To compute the expected value of an inverse function, we consider the following lemma.

*Lemma B.1*: Let $\mathbf{A} = \mathbf{Z}^H \mathbf{Z}$ be a non-singular Hermitian matrix of size $K \times K$ and $\mathbf{B} = \mathbf{A}^{-1}$, where $\mathbf{Z} \in \mathcal{C}^{N \times K}$ and $K \leq N$. We partition $\mathbf{A}$ and $\mathbf{B}$ as

$$\mathbf{A} = \begin{bmatrix} a_{11} & \mathbf{a}_{21}^H \\ \mathbf{a}_{21} & \mathbf{A}_{22} \end{bmatrix}, \quad \mathbf{B} = \begin{bmatrix} b_{11} & \mathbf{b}_{12} \\ \mathbf{b}_{21} & \mathbf{B}_{22} \end{bmatrix} \tag{46}$$

where $a_{11}(b_{11})$ is a scalar value, and the remaining terms are appropriate dimension vectors or matrices. If $a_{11} \neq 0$ and $\mathbf{A}_{22}$ is non-singular, we can express $b_{11}$ as

$$\frac{1}{b_{11}} = a_{11} - \mathbf{a}_{21}^H \mathbf{A}_{22}^{-1} \mathbf{a}_{21}. \tag{47}$$

And, if each element of $\mathbf{Z}$ is taken from i.i.d ZMCSCG random variable with variance 1, then

$$\frac{1}{b_{11}} \sim \chi_{N-K+1}^2. \tag{48}$$

*Proof:* Equation (47) can be proved by applying the well known Schur Complement theorem. The detailed derivation can also be found from *Theorem A5.2* of [54].

To prove (48) we note that both $a_{11}$ and $\mathbf{a}_{21}^H \mathbf{A}_{22}^{-1} \mathbf{a}_{21}$ are strictly non-negative real values. And when $a_{11} - \mathbf{a}_{21}^H \mathbf{A}_{22}^{-1} \mathbf{a}_{21}$ is a non-negative real valued term, by applying *Theorem 3.2.10* of [54], the probability density function of $\frac{1}{b_{11}}$ can be expressed as $\mathcal{W}_1(N - K + 1, 1)$, where $\mathcal{W}_{(.)}(.,.)$ denotes a real valued Wishart distribution. It follows

$$\frac{1}{b_{11}} \sim \mathcal{W}_1(N - K + 1, 1) \sim \chi_{N-K+1}^2 \tag{49}$$

where the second distribution is due to the fact that $\mathcal{W}_1(N - K + 1, 1)$ has the same distribution as that of Chi-square ($\chi^2$) distribution with $N - K + 1$ degrees of freedom (see *Corollary 3.2.2* of [54]). As expected when $N = K$, $\frac{1}{b_{11}}$ is a $\chi^2$ distribution with 1 degree of freedom. By setting $\mathbf{Z} = \bar{\mathbf{H}}_I$ of (45), we will get $\gamma_i = \frac{N-L+1}{\sigma^2}$.

As discussed in Section IV, the proposed design transmits $L$ symbols in $3L - 2$ sampling periods. On the other hand, the design splits the overall bandwidth $B$ into $L$ sub-bands. For these reasons, we may need to incorporate $\alpha = (\frac{3L-2}{L})(\frac{1}{L})$ to ensure that the proposed design uses the same transmit power as that of OFDM, while simultaneously transmitting $\frac{1}{\alpha}$ more symbols compared to OFDM. Upon doing so, the proposed design can achieve $R^{Pr}$ of (32).

Next we show (33) of Theorem 1. If $\gamma_0 = \frac{N}{\sigma^2} \gg 1$ and $N \gg L = 3$, we will have

$$R^{Ex} = \log_2(\gamma_0) + \log_2(L), \quad R^{Pr} = \frac{L^2}{3L-2}(\log_2(\alpha) + \log_2(\gamma_0))$$

$$R^{Pr} - R^{Ex} \geq 0 \Rightarrow \left(\frac{L^2}{3L-2} - 1\right)\log_2(\gamma_0) \geq \log_2(L) - \frac{L^2}{3L-2}\log_2(\alpha)$$

$$\Rightarrow N \geq \sigma^2 e^{(\frac{1}{L^2-3L+2}((3L-2)\log_2(L)-L^2\log_2(\alpha)))} = \max\left\{L, \sigma^2 e^{[\frac{1}{L^2-3L+2}((3L-2)\log_2(L)-L^2\log_2(\alpha))]}\right\} \quad (50)$$

∎

where the last inequality follows from the fact that $N \geq L$.

## APPENDIX B

## PROOF OF LEMMA 1

For the existing OFDM transmission, one can utilize the result of Appendix A. For the proposed approach, when $N \to \infty$, we obtain $(\mathbf{Q}^{1/2}\bar{\mathbf{H}}_I)^H(\mathbf{Q}^{1/2}\bar{\mathbf{H}}_I) \approx \text{tr}(\mathbf{Q})\mathbf{I} = N\mathbf{I}$ almost surely [19]. It follows that

$$((\bar{\bar{\mathbf{H}}}\mathbf{A})^H(\bar{\bar{\mathbf{H}}}\mathbf{A}))^{-1} \approx N^{-1}(\mathbf{A}^H\mathbf{R}\mathbf{A})^{-1} = N^{-1}\mathbf{E}, \quad (\bar{\bar{\mathbf{H}}}^H\bar{\bar{\mathbf{H}}})^{-1} \approx N^{-1}\mathbf{C} \quad (51)$$

where $\mathbf{C} = \mathbf{R}^{-1}$ and $\mathbf{E} = (\mathbf{A}^H\mathbf{R}\mathbf{A})^{-1}$ which are deterministic matrices. Thus, for the ZF detection, we will achieve

$$\gamma_i = \frac{1}{\sigma^2 \mathrm{E}\{([(\bar{\bar{\mathbf{H}}}\mathbf{A})^H(\bar{\bar{\mathbf{H}}}\mathbf{A})]^{-1})_{i,i}\}} \approx \frac{N}{\sigma^2 \mathbf{E}_{(i,i)}}. \quad (52)$$

And for the ZF-SIC detection approach, we will have

$$\gamma_i = \frac{1}{\sigma^2 \mathrm{E}\{([\bar{\bar{\mathbf{H}}}^H\bar{\bar{\mathbf{H}}}]^{-1})_{S,S}\}} \approx \frac{N}{\sigma^2 \mathbf{C}_{(i,i)}}. \quad (53)$$

## REFERENCES

[1] E. Hossain, M. Rasti, H. Tabassum, and A. Abdelnasser, "Evolution towards 5G multi-tier cellular wireless networks: An interference management perspective," *IEEE Wireless Commun. Mag.*, pp. 118 – 127, Jun. 2014.

[2] A. Osseiran, F. Boccardi, V. Braun, K. Kusume, P. Marsch, M. Maternia, O. Queseth, M. Schellmann, H. Schotten, H. Taoka, H. Tullberg, M. A. Uusitalo, B. Timus, and M. Fallgren, "Scenarios for 5G mobile and wireless communications: The vision of the METIS project," *IEEE Commun. Mag.*, pp. 26 – 35, May 2014.

[3] S. Sugiura, "Frequency-domain equalization of faster-than-Nyquist signaling," *IEEE Wireless Commun. Letters*, vol. 2, no. 5, pp. 555 – 558, Oct. 2013.

[4] J. B. Anderson, F. Rusek, and V. Owall, "Faster-than-Nyquist signaling," *IEEE Proceedings.*, pp. 1817 – 1830, Aug. 2013.

[5] T. Bogale and L. Le, "Massive MIMO and mmWave for 5G wireless HetNet: Potential benefits and challenges," *IEEE Vehic. Techno. Mag.*, vol. 11, no. 1, pp. 64 – 75, Mar. 2016.

[6] L. B. Le, V. Lau, E. Jorswieck, N.-D. Dao, A. Haghighat, D.-I. Kim, and T. Le-Ngoc, "Enabling 5G mobile wireless technologies," *EURASIP J. Wireless Commun. Networking.*, vol. 218, pp. 1 – 14, 2015.

[7] J. E. Mazo., "Faster-than-Nyquist signaling," *Bell Syst. Technical J.*, vol. 54, no. 8, pp. 1451 – 1462, Oct. 1975.

[8] A. D. Liveris and C. N. Georghiades, "Exploiting faster-than-Nyquist signaling," *IEEE Trans. Commun.*, vol. 51, no. 9, pp. 1502 – 1511, Sep. 2003.

[9] M. E. Hefnawy and H. Taoka, "Overview of faster-than-Nyquist for future mobile communication systems," in *Proc. IEEE Veh. Technol. Conf (VTC-Spring).*, 2013, pp. 1 – 5.

[10] M. E. Hefnawy and G. Kramer, "Impact of spectrum sharing on the efficiency of faster-than-Nyquist signaling," in *Proc. IEEE Wireless Commun. and Networking. Conf.*, 2014, pp. 648 – 653.

[11] M-S. Baek, N-H. Hur, and H. Lim, "Novel interference cancellation technique based on matrix computation for FTN communication system," in *Proc. IEEE Military Commun. Conf.*, 2014, pp. 830 – 834.

[12] F. Schaich and T. Wild, "A reduced complexity receiver for multi-carrier faster-than-Nyquist signaling," in *Proc. IEEE Global Telecommun. Workshop*, 2013, pp. 235 – 240.

[13] A. Prlja and J. B. Anderson, "Reduced-complexity receivers for strongly narrowband intersymbol interference introduced by faster-than-Nyquist signaling," *IEEE Trans. Commun.*, vol. 60, no. 9, pp. 2591 – 2601, Sep. 2012.

[14] F. Rusek, "On the existence of the Mazo-limit on MIMO channels," *IEEE Trans. Wireless Commun.*, vol. 8, no. 3, pp. 1118 – 1121, Mar. 2009.

[15] F. Rusek, "A first encounter with faster-than-Nyquist signaling on the MIMO channel," in *Proc. IEEE Wireless Commun. and Networking. Conf.*, 2007, pp. 1094 – 1098.

[16] L. Zheng and D. N. C. Tse, "Diversity and multiplexing: A fundamental tradeoff in multiple-antenna channels," *Trans. Commun.*, vol. 49, no. 5, pp. 1073 – 1096, 2003.

[17] D. Tse and P. Viswanath, *Fundamentals of Wireless Communication*, Cambridge University Press, 2005.

[18] D. P. Palomar, *A unified framework for communications through MIMO channels*, Ph.D. thesis, Technical University of Catalonia (UPC), Barcelona, Spain, 2003.

[19] H. Q. Ngo, E. G. Larsson, and T. L. Marzetta, "Energy and spectral efficiency of very large multiuser MIMO systems," *IEEE Trans. Commun.*, vol. 61, no. 4, pp. 1436 – 1449, Apr. 2013.

[20] J. Hoydis, C. Hoek, T. Wild, and S. T. Brink, "Channel measurements for large antenna arrays," in *Proc. IEEE International Symposium Wireless Communication Systems (ISWCS)*, 2012, pp. 811 – 815.

[21] X. Gao, F. Tufvesson, O. Edfors, and F. Rusek, "Measured propagation characteristics for very-large MIMO at 2.6 GHz," in *Proc. Asilomar Conf. Sign. Syst. and Comp.*, 2012, pp. 295 – 299.

[22] X. Gao, O. Edfors, F. Rusek, and F. Tufvesson, "Massive MIMO performance evaluation based on measured propagation data," *IEEE Trans. Wireless Commun.*, vol. 14, no. 7, pp. 3899 – 3911, Jul. 2015.

[23] X. Gao, O. Edfors, F. Rusek, and F. Tufvesson, "Linear pre-coding performance in measured very-large MIMO channels," in *Proc. IEEE Veh. Technol. Conf (VTC-Fall).*, 2011, pp. 1 – 5.

[24] A. Manolakos, M. Chowdhury, and A. Goldsmith, "Energy-based modulation for noncoherent massive SIMO systems," Jul. 2015, http://arxiv.org/abs/1507.04978.

[25] "Jensen's inequality," www.econ.hku.hk/~wsuen/teaching/micro/jensen.pdf.

[26] T. E. Bogale, *Transceiver design for single-cell and multi-cell downlink multiuser MIMO systems*, Ph.D. thesis, University Catholique de Louvain (UCL), Louvain la neuve, Belgium, 2013.

[27] H. Zarrinkoub, *Understanding LTE with MATLAB: From Mathematical Modeling to Simulation and Prototyping*, John W&S, Ltd, 2014.

[28] J. Zyren, "Overview of the 3GPP long term evolution physical layer," in *Freescale semiconductor*, 2007, pp. 1 – 24.

[29] T. E. Bogale, L. B. Le, X. Wang, and L. Vandendorpe, "Pilot contamination in wideband massive MIMO system: Number of cells vs multipath," in *Proc. IEEE Global Telecommun. Conf.*, 2015.

[30] A. Goldsmith, *Wireless Communications*, Cambridge University Press, New York, NY, USA, 2005.

[31] Y. Liu, Z. Tan, H. Hu, L. J. Cimini, and G. Y. Li, "Channel estimation for OFDM," *IEEE Commun. Surv. Tutor.*, vol. 16, no. 4, pp. 1891 – 1908, 2014.

[32] 3GPP TS 36.104, "Base station (BS) radio transmission and reception," 2012, 3rd Generation Partnership Project; Technical Specification Group Radio Access Network; Evolved Universal Terrestrial Radio Access (E-UTRA).

[33] ETSI TR 125 943, "Universal mobile telecommunications system (UMTS); deployment aspects," 2010, 3GPP TR 25.943 version 9.0.0 Release 9.

[34] T. L. Marzetta, "Noncooperative cellular wireless with unlimited numbers of base station antennas," *IEEE Trans. Wireless Commun.*, vol. 9, no. 11, pp. 3590 – 3600, Nov 2010.

[35] J. G. Andrews, A. Ghosh, and R. Muhamed, *Fundamentals of WiMAX: Understanding Broadband Wireless Networking*, Pearson Edu., 2007.

[36] S.M. Alamouti, "A simple transmit diversity technique for wireless communications," *IEEE J. Select. Areas in Commun.*, vol. 16, no. 8, pp. 1451 – 1458, 1998.

[37] F. Rusek and J. B. Anderson, "On information rates for faster than Nyquist signaling," in *Proc. IEEE Global Telecommun. Conf.*, 2006, pp. 1 – 5.

[38] V. Venkateswaran and A-J. V. Veen, "Analog beamforming in MIMO communications with phase shift networks and online channel estimation," *IEEE Trans. Signal Process.*, vol. 58, no. 8, pp. 4131 – 4143, Aug. 2010.

[39] X. Zhang, A. F. Molisch, and S-Y. Kung, "Variable-phase-shift-based RF-Baseband codesign for MIMO antenna selection," *IEEE Trans. Signal Process.*, vol. 53, no. 11, pp. 4091 – 4103, Nov. 2005.

[40] T. E. Bogale, L. B. Le, A. Haghighat, and L. Vandendorpe, "On the number of RF chains and phase shifters, and scheduling design with hybrid analog-digital beamforming," *IEEE Trans. Wireless Commun.*, vol. 15, no. 5, pp. 3311–3326, May 2016.

[41] Francois Le Gall, "Powers of tensors and fast matrix multiplication," in *Proceedings of the 39th International Symposium on Symbolic and Algebraic Computation*, New York, NY, USA, 2014, ISSAC, pp. 296–303, ACM.

[42] T. Moazzeni, "On the compactness of ofdm and hermite signals," *IEEE Commun. Letters*, vol. 20, no. 7, pp. 1313–1316, July 2016.

[43] A. J. Goldsmith and S-G. Chua, "Variable-rate variable-power MQAM for fading channels," *IEEE Trans. Commun.*, vol. 45, no. 10, pp. 1218 – 1230, Oct. 1997.

[44] A. J. Goldsmith and P. P. Varaiya, "Capacity of fading channels with channel side information," *IEEE Trans. Info. Theory*, vol. 43, no. 6, pp. 1986 – 1992, Nov. 1997.

[45] S. L. Loyka, "Channel capacity of MIMO architecture using the exponential correlation matrix," *IEEE Commun. Letters*, vol. 5, no. 9, pp. 369 – 371, 2001.

[46] M. Ding, *Multiple-input multiple-output wireless system designs with imperfect channel knowledge*, Ph.D. thesis, Queens University Kingston, Ontario, Canada, 2008.

[47] M. Patzold, A. Szczepanski, and N. Youssef, "Methods for modeling of specified and measured multipath power-delay profiles," *IEEE Trans. Veh. Technol.*, vol. 51, no. 5, pp. 978 – 988, Sep. 2002.

[48] D. R. Morgan, "Analysis and realization of an exponentially-decaying impulse response model for frequency-selective fading channels," *IEEE Signal Process. Letters*, , no. 15, pp. 441 – 444, Sep. 2008.

[49] T. E. Bogale, L. B. Le, and X. Wang, "Hybrid analog-digital channel estimation and beamforming: Training-throughput tradeoff," *IEEE Trans. Commun.*, Dec. 2015.

[50] T. E. Bogale and L. Vandendorpe, "Weighted sum rate optimization for downlink multiuser MIMO coordinated base station systems: Centralized and distributed algorithms," *IEEE Trans. Signal Process.*, vol. 60, no. 4, pp. 1876 – 1889, Dec. 2011.

[51] S. Vishwanath, N. Jindal, and A. Goldsmith, "Duality, achievable rates, and sum-rate capacity of Gaussian MIMO broadcast channels," *IEEE Trans. Info. Theory*, vol. 49, no. 10, pp. 2658 – 2668, Oct. 2003.

[52] P. Viswanath and D. Tse, "Sum capacity of the vector Gaussian broadcast channel and uplink-downlink duality," *IEEE Trans. Info. Theory*, vol. 49, no. 8, pp. 1912 – 1921, Aug. 2003.

[53] T. E. Bogale and L. Vandendorpe, "Linear transceiver design for downlink multiuser MIMO systems: Downlink-interference duality approach," *IEEE Trans. Signal Process.*, vol. 61, no. 19, pp. 4686 – 4700, Oct. 2013.

[54] R. J. Muirhead, *Aspects of Multivariate Statistical Theory*, John Wiley and Sons Inc., New Jersey, 1982.